\documentclass[ALICE,manyauthors]{cernphprep}

\newcommand{\pp}{p\kern-0.05em p}
\newcommand{\ppbar}{\mathrm{p}\kern-0.05em \overline{\mathrm{p}}}

\newcommand{\GeV}{\ensuremath{\mathrm{GeV}\kern-0.05em}}
\newcommand{\GeVc}{\ensuremath{\mathrm{GeV}\kern-0.05em/\kern-0.02em c}}

\newcommand{\pT}{\ensuremath{p_{\mathrm{T}}}}
\newcommand{\RL}{\ensuremath{R_{\mathrm{L}}}}

\newcommand{\pTtrack}{\ensuremath{p_{\mathrm{T,track}}}}
\newcommand{\sigmaeec}{\ensuremath{\Sigma_{\mathrm{EEC}}}}

\newcommand{\kT}{\ensuremath{k_{\mathrm{T}}}}

\newcommand{\etajet}{\ensuremath{\eta_{\mathrm{jet}}}}
\newcommand{\pTjet}{\ensuremath{p_{\mathrm{T}}^{\mathrm{jet}}}}
\newcommand{\pTchjet}{\ensuremath{p_{\mathrm{T}}^{\mathrm{ch\; jet}}}}

\usepackage[comma,square,numbers,sort&compress]{natbib}
\usepackage{hyperref}
\usepackage{lineno}
\usepackage{amsmath}
\usepackage{bigints}  
\usepackage{afterpage}  
\usepackage{multirow}
\usepackage{xcolor}
\usepackage[Symbol]{upgreek}
\usepackage[T1]{fontenc}
\usepackage{mathtools}
\usepackage{siunitx}
\usepackage{soul}
\usepackage{orcidlink}
\usepackage{comment}

\DeclarePairedDelimiterXPP\BigOSI[2]%
  {\mathcal{O}}{(}{)}{}%
  {\SI{#1}{#2}}

\begin{document}%

\begin{titlepage}
\PHyear{2024}
\PHnumber{245}
\PHdate{16 Sep}
%

\title{Exposing the parton-hadron transition within jets with energy-energy correlators in \pp{} collisions at $\mathbf{\sqrt{\textit s}=5.02}  \ \mathbf{TeV}$}

\ShortTitle{Measurement of the energy-energy correlators in \pp{} collisions at $\sqrt{s} = 5.02$ TeV}   

\Collaboration{ALICE Collaboration\thanks{See Appendix~\ref{app:collab} for the list of collaboration members}}
\ShortAuthor{ALICE Collaboration} 

\begin{abstract}
This paper presents a fully-corrected measurement of the energy-energy correlator (EEC) within jets in \pp{} collisions. The EEC traces the energy flow as a highly energetic parton undergoes a QCD shower followed by the confinement of partons into hadrons, probing the correlation function of the energy flow inside jets. 
The EEC observable is measured as a function of the charged particle pair angular distance, \RL, for 20 < \pTchjet~< 80 \ \GeVc. In the perturbative region (large \RL), a good agreement between the data and a next-to-leading-log perturbative QCD calculation is observed. In the non-perturbative region (small \RL), the data exhibits a linear \RL~dependence. There is a transition region in between, characterized by a turnover in the EEC distribution, indicating where hadronization effects begin to dominate. The peak of this transition region is located at $2.43 \pm 0.14~\GeVc/\langle\pTchjet\rangle$ for jets of various energies, indicating a common energy scale for the hadronization process.
State-of-the-art Monte Carlo event generators are compared with the measurements, and can be used to constrain the parton shower and hadronization mechanisms. 

\end{abstract}
\end{titlepage}

\setcounter{page}{2}

\section{Introduction}

Jets serve as powerful probes for studying quantum chromodynamics (QCD). At the Large Hadron Collider (LHC), they are produced abundantly via hard scattering processes (large momentum transfer) between quarks and gluons (partons) from each colliding proton. The high-energy partons from the initial scattering undergo QCD radiation and split into lower energy partons. This process can be described using perturbative QCD (pQCD) calculations and modeled by Monte Carlo (MC) parton shower routines. Below a certain energy scale, 
color-neutral particles emerge when partons are confined into hadrons; eventually the final stable particles stream into detectors. The internal structure of the particles inside jets, referred to as jet substructure, encodes both the radiation pattern of the parton shower and information about the hadronization process~\cite{Salam:2010nqg}. To unravel the QCD dynamics within the fragmentation and hadronization processes, a variety of jet substructure observables have been studied ~\cite{ALICE:2014dla,ALICE:2018gni,ALICE:2018ype,ALICE:2019cbr,ALICE:2019ykw,ALICE:2021njq,ALICE:2021obz,ALICE:2021vrw,ALICE:2021aqk,ALICE:2022rdg,ALICE:2022vsz, ATLAS:2012am,ATLAS:2012nnf,ATLAS:2013uet,ATLAS:2017zda,ATLAS:2019kwg,ATLAS:2019rqw,ATLAS:2019mgf,ATLAS:2020bbn,ATLAS:2022vii,Aad:2019wdr, CMS:2012oyn,CMS:2012nro,CMS:2014jjt,CMS:2017yer,CMS:2017qlm,CMS:2018vzn,CMS:2018jco,CMS:2021iwu,CMS:2020geg,LHCb:2017llq,LHCb:2019qoc, STAR:2020ejj,STAR:2021lvw,STAR:2021kjt}. Measurements of the jet substructure offer new opportunities for studying the properties of nuclei and the quark--gluon plasma (QGP) via the interplay between the jet and the cold or the hot QCD medium in proton--nucleus (pA) and nucleus--nucleus (AA) collisions, respectively~\cite{Connors:2017ptx, ALICE:2022wpn, Albacete:2017qng}.
 
A novel jet substructure observable, referred to as energy-energy correlator (EEC)~\cite{Chen:2020vvp, Komiske:2022enw, Lee:2022ige}, focuses on the correlation function of energy flow inside jets. Originally introduced as an event shape observable in $e^{+}e^{-}$ collisions~\cite{Basham:1978bw,ACTON1992547,OPAL:1993pnw,Kardos:2018kqj} or as the observable needed to construct jet discriminators~\cite{Grigoriev:2003yc, Sveshnikov:1995vi, Grigoriev:2003tn}, the EEC is infrared-collinear (IRC) safe~\cite{Tkachov:1999py} and can be calculated from first principles in QCD in the perturbative limit~\cite{Kardos:2018kqj}. 
It has been used to constrain the strong coupling constant $\alpha_s$~\cite{ACTON1992547,OPAL:1993pnw,Kardos:2018kqj}. EEC observables can be defined in different ways, offering precision probes of both perturbative and non-perturbative QCD dynamics in collisions ranging from $e^{+}e^{-}$ annihilation and hadronic collisions, to deep inelastic scattering (DIS), as summarized in Ref.~\cite{Neill:2022lqx, Moult:2025nhu}. 

\begin{figure}
    \centering
    \includegraphics[width=0.5\textwidth, height=0.3\textwidth]{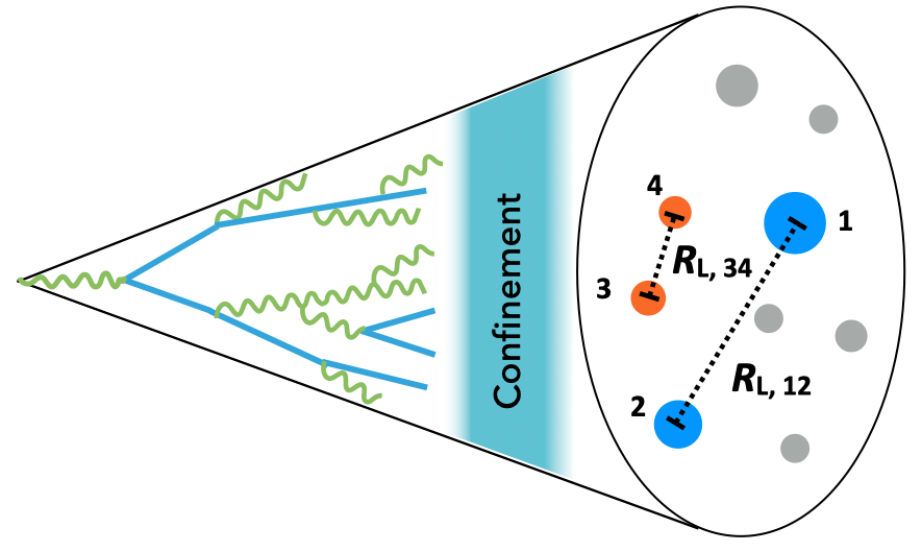}
    \caption{A schematic of the jet evolution through parton shower (lines), confinement process (shaded area), to final state particles (circles).}
    \label{fig:EEC-schematic}
\end{figure}

Experimentally, the EEC is an energy-weighted two-particle correlation as a function of the angular distance between pairs of particles.
In this paper, we study the EEC inside high-energy jets produced in \pp{} collisions at $\sqrt{s} =5.02$~TeV. This 
approach focuses on the energy correlation function in the collinear limit instead of over the whole event, 
providing a new perspective on jet evolution, from perturbative splittings to parton confinement into hadrons~\cite{Chen:2020vvp}. The energy-energy correlation function, $\Sigma_{\text{EEC}}(R_{\text{L}})$, is defined as the following:

\begin{equation}
    \Sigma_{\text{EEC}}(R_{\text{L}}) = \frac{1}{ N_{\text{jet}}\cdot\Delta}\int_{R_{\text{L}}-\frac{1}{2}\Delta}^{R_{\text{L}}+\frac{1}{2}\Delta}\sum_{\text{jets}}\sum_{i, j}\frac{p_{\text{T}, i}p_{\text{T}, j}}{(\pTjet)^{2}}\delta(R_{\text{L}}'-R_{\text{L}, ij})dR_{\text{L}}'.
\end{equation}
The sum runs over all final state particle pairs ($i$, $j$) inside each jet \footnote{Both pair ($i$, $j$) and ($j$, $i$) are considered. For example, a pair of particle 1 and particle 2 is counted twice, (1, 2) and (2, 1).}. The angular distance between each pair in the $\eta-\varphi$ plane is $R_{\text{L}, ij} = \sqrt{(\varphi_j - \varphi_i)^2 + (\eta_j - \eta_i)^2}$. $\Delta$ is the angular bin width and $N_{\text{jet}}$ is the total number of jets.
Fig.~\ref{fig:EEC-schematic} illustrates how particles inside a jet are paired when constructing the EEC observable.

Recent theoretical developments have enabled the analytical calculation of energy correlators inside jets to high accuracy~\cite{Chen:2020vvp}, and the calculation has been extended to charged particles using track functions~\cite{Chen:2022pdu,Chen:2022muj,Jaarsma:2022kdd,Li:2021zcf, Lee:2022ige}. 
As the soft radiation is suppressed by the energy weight, 
this observable has reduced sensitivity to higher-order corrections, the underlying event, and soft jet fragments in the perturbative regime.

Another advantage of the EEC observable arises from the angular scale, \RL. 
Based on the angular ordering of the QCD radiation (splitting), the time scale $\tau$ of splitting can be approximated by $\tau \sim 1/(\pTjet\cdot \RL^{2})$, where $\tau$ is the splitting time and \pTjet~represents the scale of the initial hard process~\cite{Chen:2020vvp, Komiske:2022enw}.
The large-\RL~regime corresponds to the perturbative regime during the early parton splitting stage of the jet dynamics, hence referred to as the perturbative or parton scaling region. 
Non-perturbative effects become more significant as the QCD radiation evolves towards lower energy scales and smaller angles. Below a certain angular scale, confinement effects are expected to
dominate. This separation of the perturbative and non-perturbative regimes into distinct angular regions was already observed by the Compact Muon 
Solenoid (CMS) experiment~\cite{Komiske:2022enw, CMS:2024mlf}, and they have extracted $\alpha_{s}$ to high precision using the energy correlators in the perturbative regime.  

In the paper we take advantage of the excellent track reconstruction capabilities of the A Large Ion Collider Experiment (ALICE)~\cite{ALICE:2022wpn} to construct the EEC using charged particle tracks within reconstructed charged-particle jets. 
The measurement of jet EEC in \pp~collisions with detector effects fully corrected at the LHC is presented. 
The measurements are reported in charged particle jet \pT~(\pTchjet) intervals of 20--40, 40--60, 60--80~\GeVc, which marks the first EEC measurement in jets in such a low-\pT~kinematic regime at the LHC. In this regime, the measured \RL~range extends from the perturbative to the non-perturbative regime, allowing a closer examination of the transition region between them. The results are compared with state-of-the-art MC models with different hadronization mechanisms and pQCD calculations. The transition region is analyzed to study hadronization as a function of jet \pT.

\section{Experimental setup and data sets}
\label{sec:exp_data}

This analysis utilizes  \pp~collisions at $\sqrt{s} = 5.02$ TeV collected by the ALICE experiment in 2017. Minimum bias events are triggered by coincidence hits on the V0 scintillator detectors~\cite{ALICE:2013axi} at forward and backward rapidity. Events are required to have a primary vertex within $\pm 10$ cm of the nominal interaction point to provide uniform acceptance and high tracking efficiency. Events with more than one reconstructed collision vertex are removed to avoid pileup of multiple collisions. A total of 870 million minimum bias events passed the event selection criteria, corresponding to an integrated luminosity of 18.0 $\pm$ 0.4 nb$^{-1}$~\cite{ppXsec}.

The central barrel tracking system is the main detector used in this analysis. The central tracker consists of the Inner Tracking System (ITS)~\cite{ALICE:2010tia} and Time Projection Chamber (TPC)~\cite{Alme:2010ke}. They are inside a 0.5 T magnetic field and have a high efficiency for detection of charged particles with \pT~(\pTtrack) from 0.15 to 100~\GeVc, $|\eta|<0.9$, and $0 < \varphi < 2\uppi$. The single track efficiency is \pT~dependent, increasing from $\sim60\%$ at 0.15~GeV to $\sim80\%$ at 1~GeV and remaining above 75$\%$ at higher \pT. The momentum resolution, $\sigma_p/p$, is $\approx 1\%$ at $\pTtrack=1$ \GeVc~increasing to $\approx 4\%$ at $\pTtrack=40$ \GeVc. The angular resolution is $\approx$1 mrad at $\pTtrack=1$ \GeVc~decreasing to < 0.6 mrad for $\pTtrack>5$~\GeVc. ALICE uses a standard cylindrical coordinate system where $\eta$ is the pseudorapidity and $\varphi$ is the azimuthal angle. More detailed descriptions of the ALICE detector and performance can be found in Ref.~\cite{ALICE:2014sbx,ALICE:2008ngc}.

\section{Analysis method}
\label{sec:analysis}

Charged particles measured by the central barrel tracking system are selected using the same criteria as in Ref~\cite{ALICE:2013dpt, ALICE:2014sbx}. The selected charged particles are clustered into jets via the anti-\kT~algorithm with $R=0.4$ using the FastJet 3.3.3 package~\cite{Cacciari:2011ma}. The E-scheme recombination is used, assigning all charged particles the pion mass. Reconstructed jets with $|\etajet|<0.9-R$ are used to avoid detector edge effects in the jet reconstruction. The EEC observable is measured in charged jet $p_T$ ranges of 20 < \pTchjet~< 40~\GeVc, 40 < \pTchjet~< 60~\GeVc, and 60 < \pTchjet~< 80~\GeVc. 
Only constituents inside the reconstructed jets with $\pTtrack > 1 \ \GeVc$ enter the EEC, in order to reduce the impact of non-jet particles from the underlying event (UE) \footnote{In PYTHIA8, the $\pTtrack > 1 \ \GeVc$ cut reduces the multi-parton interactions (MPI) effect from up to 6$\%$ to up to 3$\%$ in the large $R_L$ region.}. No additional corrections are applied for the presence of the UE. 

Detector effects are assessed by simulating pp collisions with PYTHIA 8 (version 8.210)~\cite{Sjostrand:2014zea} Monash tune~\cite{Skands:2014pea} and then propagating the final-state particles through a GEANT3 model~\cite{Brun:1119728} of the ALICE detector. The key detector effects for EEC are the track momentum resolution, angular resolution, and efficiency for both single tracks and track pairs. The track momentum resolution and efficiency affect the calculated energy weights, the number of reconstructed pairs, and the reconstructed \pTchjet. Track pairs at very small angular separation with the same charge have very similar trajectories in the tracking volume, which may cause the track reconstruction algorithm to miss one of the tracks or mis-reconstruct them. This effect is referred to as the track-merging effect, and was quantified for the femtoscopy measurements in ALICE~\cite{ALICE:2018fdu}. Track merging decreases the pair efficiency for tracks at small
distance, causing a reduced $\Sigma_{\text{EEC}}$ at low \RL. A $\sim90\%$ pair efficiency is observed in the GEANT simulation for pairs with $\RL > 0.01$, which drops steeply to $<20\%$ for pairs with smaller $\RL$. To achieve precise measurements with small systematic uncertainties, results are reported for $\RL > 0.01$. 

The values of the corrections are $<10\%$, except for the \RL~bins near 0.01, for which the corrections are about $20\%$.
As each track pair has a unique energy weight, full unfolding to correct for detector effects is complicated and requires higher statistics. However, the excellent momentum and position resolution in ALICE results in little bin migration in \RL. Hence, in this analysis, instead of unfolding, a two-dimensional bin-by-bin correction is utilized where the correction factors are obtained from the simulation by
comparing the generator-level and corresponding detector-level information for each \pTchjet{} and \RL~bin. The bin-by-bin correction can be more sensitive to differences between the generator used for deriving the correction factor and the physics events collected; this is included in the systematic uncertainty. 
A cross-check was performed using a Bayesian unfolding technique~\cite{DAgostini, Adye:2011gm}. The difference between the unfolded and bin-by-bin corrected EEC distributions is small and is included in the systematic uncertainties. More information about the unfolding procedure can be found in Appendix~\ref{app:new}.

\section{Systematic uncertainties}
\label{sec:systematics}

The main systematic uncertainty sources in this measurement are the single-track efficiency, the track pair efficiency, the generator dependence of the correction factors and the unfolding cross check. 
The single-track efficiency uncertainty is estimated to be $\approx 3\%$ based on previous studies where the track selection and ITS-TPC matching criteria are varied~\cite{ALICE:2019wqv}. 
The impact of this uncertainty is determined by randomly rejecting $3\%$ of the detector-level tracks in simulated events and re-calculating the correction factors. It leads to a \RL~dependent change of $< 3\%$, which is quoted as the systematic uncertainty for single-track efficiency and also includes the track efficiency uncertainty effect upon the jet \pT. The systematic uncertainty due to the pair efficiency is evaluated by studying the variation of the corrected data points with different pair selection parameters, yielding $< 2\%$ variation in the \RL~range reported. The uncertainty from the generator dependence of the correction factors is determined by comparing the correction factors extracted from events simulated with PYTHIA 8~\cite{ALICE:2019wqv} and Herwig 7~\cite{Bellm:2015jjp}, with parameterized detector responses included, yielding $< 3\%$ variation. The unfolded EEC shows $< 3\%$ difference from the bin-by-bin corrected results, and the difference is assigned as a systematic uncertainty on the bin-by-bin correction method. The systematic uncertainties are assumed to be independent and are added in quadrature, resulting in a total uncertainty of $< 4\%$ for most of the reported \RL~range.
 
\section{Results and discussion}
\label{sec:results}

\begin{figure}
    \centering
    \includegraphics[width=0.50\textwidth]{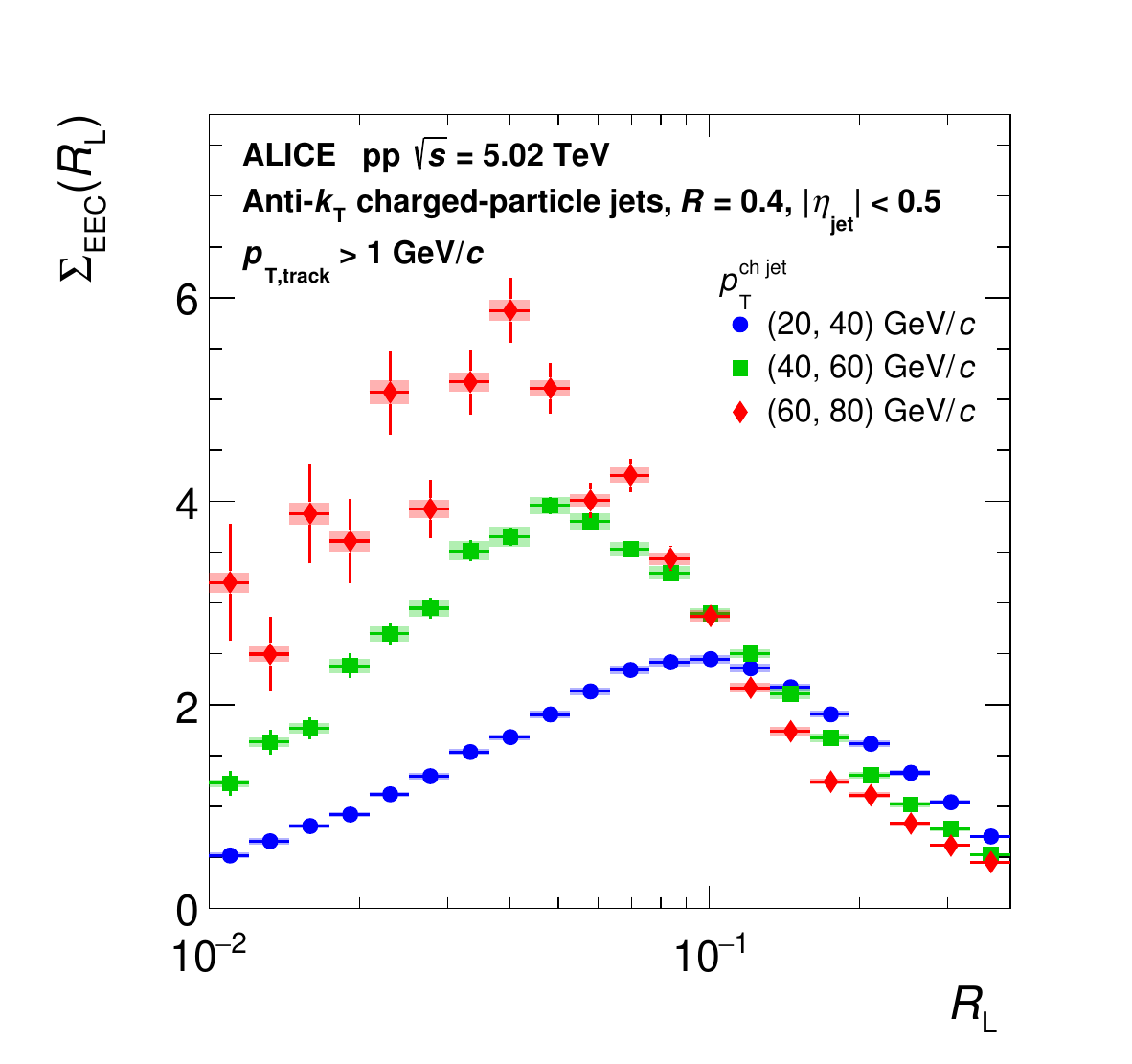}
    \caption{Fully corrected $\Sigma_{\text{EEC}}$ as a function of $R_L$ in the \pTchjet{} intervals 20--40, 40--60, and 60--80~\GeVc.}
    \label{fig:pT-compairison}
\end{figure}

The EEC distributions for \pTchjet~ranges of 20--40, 40--60, and 60--80~\GeVc~are shown in Fig.~\ref{fig:pT-compairison}. In each \pTchjet~range, distinct \RL~dependences are observed at the large-angle and small-angle limits, with a transition region in between. Similar distinct regions have been observed in EEC measurements from pp collisions at different beam energies~\cite{CMS:2024mlf, STAR:2025jut} and in $e^{+}e^{-}$ collisions~\cite{Bossi:2025xsi}. The three regions can be intuitively understood with the following explanations:
\begin{itemize}
    \item The region at large \RL~(partonic degrees of freedom) is dominated by the early time jet evolution where partons split and shower perturbatively. Harder splittings with larger \RL~occur less frequently than soft emissions with smaller \RL, resulting in \sigmaeec~scaling approximately as 1/\RL.
    \item The region at small \RL~(hadronic degrees of freedom) reflects the jet structure at later times dominated by the features of QCD inaccessible with perturbative methods, namely hadronization. Hadron pairs with larger \RL~occur more frequently as the available phase space increases, resulting in \sigmaeec~scaling approximately linearly with \RL.
    \item The intermediate \RL~transition region exhibits a peak structure where the hadronization contribution diminishes toward larger \RL~and the perturbative splitting contribution diminishes toward smaller \RL.

\end{itemize}

The EEC distributions exhibit a significant jet \pT~dependence, even within the limited range of jet momenta in this work. For jets in the higher \pTchjet~range, the perturbative region extends to smaller \RL\ values, consequently shifting the peak of the transition region to a lower \RL. This behavior is expected since the energy scale or virtuality, $\mu$, of the parton splitting is proportional to the product of the jet \pT~and \RL. When $\mu$ approaches the confinement scale $\Lambda$ from above, parton splitting ceases and the $\Sigma_{\text{EEC}}$ stops increasing. Instead, hadronization takes over and shapes the EEC distributions down to \RL~= 0. As a result, the transition peak should be located at $\RL \sim \Lambda/\pTchjet$~\cite{Dokshitzer:1999sh, Korchemsky:1999kt}, implying that the EEC in more energetic jets peaks at smaller angular scales, consistent with our observations.

\begin{figure}
    \centering
    \includegraphics[width=0.50\textwidth]{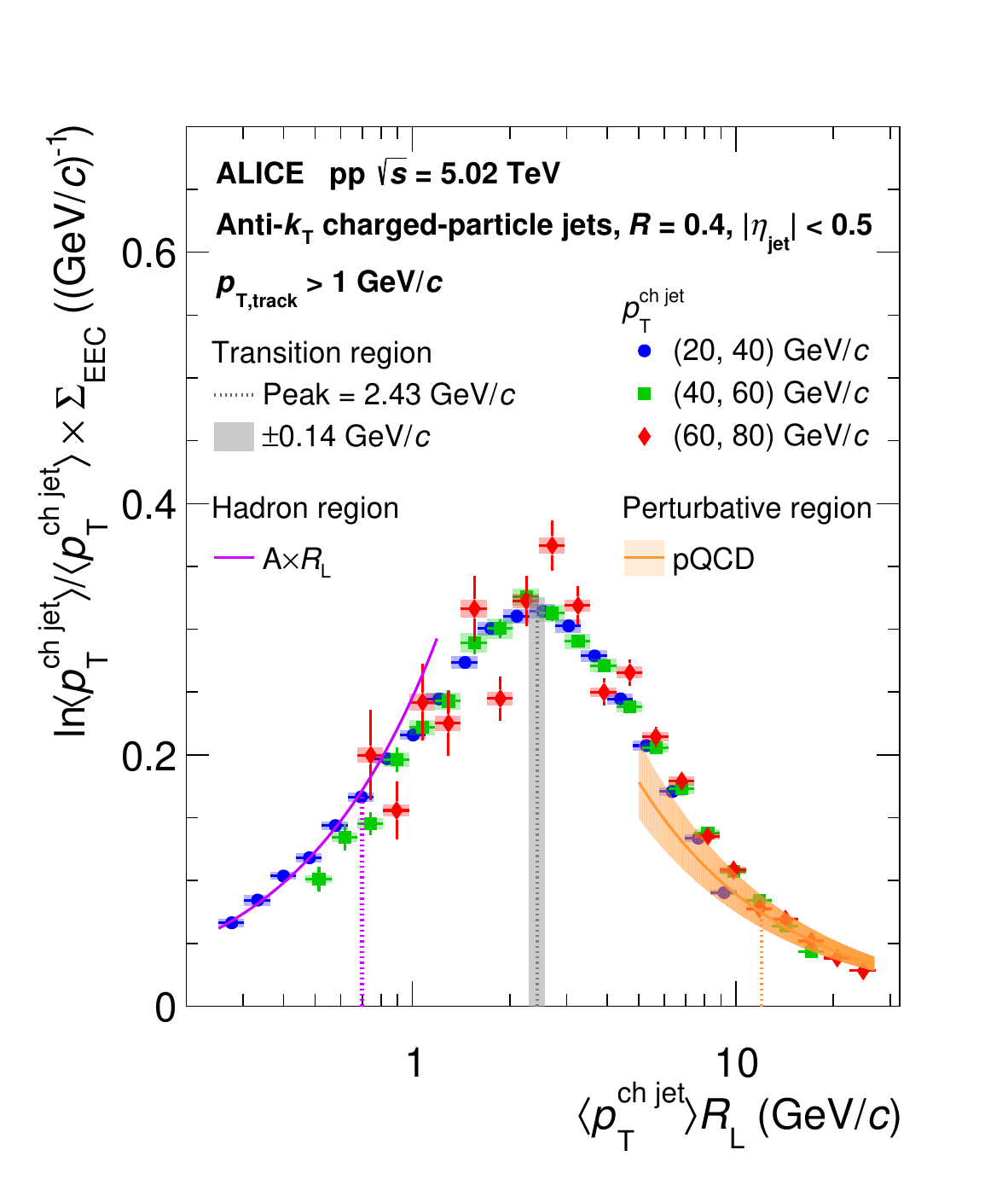}
    \caption{Normalized $\Sigma_{\text{EEC}}$ as a function of $\langle\pTchjet\rangle\RL$. $\ln\langle\pTchjet\rangle$ in the y-axis represents $\ln(\langle\pTchjet\rangle/(\GeVc))$, a unit-less number, used throughout this paper for simplicity in notation. The gray line corresponds to the maximum location of the distribution and the gray band corresponds to a $\pm$0.14 \GeVc~uncertainty along the x-axis. The orange curves show pQCD calculations ~\cite{Lee:2022ige}, which are normalized to data such that the integral inside \RL~range of 12~\GeVc/$\langle\pTchjet\rangle$, marked by the vertical dashed orange line, to 0.4 are the same. The purple curve represents a linear functional form that is fit to data ($\chi^2 = 1.65$) in the \RL~range of 0.01 to $0.7~\GeVc/\langle\pTchjet\rangle$, marked by the vertical dashed purple line. As the fitting range for the linear curve is mostly accessible by the data in 20--40 \GeVc, the comparison of linear fit to data is only shown for 20--40 \GeVc. As the normalization range for the orange pQCD curve is mostly accessible by the data in 60--80 \GeVc, the comparison of pQCD to data is only shown for 60--80 \GeVc.}
    \label{fig:pT-shifted-compairison}
\end{figure}

To further examine the dependence of the transition region on the energy of jets, the peak position ($\RL^{\text{peak}}$) and peak height ($\sigmaeec^{\text{peak}}$) are extracted for the measured \pTchjet~ranges (see Appendix~\ref{app:B}). We find that $\RL^{\text{peak}}$ is proportional to $ 1/\langle\pTchjet\rangle$ and $\sigmaeec^{\text{peak}}$ is proportional to 
$ \langle\pTchjet\rangle/\ln\langle\pTchjet\rangle$\footnote{\label{fn:lnpt} The $\langle\pTchjet\rangle$ inside  $\ln\langle\pTchjet\rangle$ actually represents $\langle\pTchjet\rangle/(\GeVc)$, which is a unit-less number. For simplicity in notation, we use $\ln\langle\pTchjet\rangle$ for $ \ln(\langle\pTchjet\rangle/(\GeVc))$ throughout this paper.}. $\langle\pTchjet\rangle$ is the average \pT~for charged-particle jets in each \pT~bin, extracted from PYTHIA 8 (Monash tune) reweighted to reproduce the ALICE data in Ref.~\cite{ALICE:2019wqv}. To directly visualize the jet \pT~dependence of the transition peak position and height, 
$\Sigma_{\text{EEC}}$ as a function of $\langle\pTchjet\rangle\RL$ is shown in Fig.~\ref{fig:pT-shifted-compairison}. The vertical scale of the EEC distribution is scaled by $\ln\langle\pTchjet\rangle/\langle\pTchjet\rangle$. Strikingly, the scaled EEC distributions, $\ln\langle\pTchjet\rangle/\langle\pTchjet\rangle\Sigma_{\text{EEC}}$, for different \pTchjet~bins collapse into a common curve. The transition peak location has been investigated in several recent theoretical studies~\cite{Liu:2024lxy, Barata:2024wsu}, where it is found to be described by non-perturbative transverse momentum dependent fragmentation functions (TMD). This framework provides predictive power for the transition position in other collision systems such as $e^+e^-$ annihilation. Importantly, both theoretical calculations and Pythia studies~\cite{ALICE:2025igw, Apolinario:2025vtx} indicate that the EEC distribution and transition position depend on the color charge and mass of the jet-initiating parton. In the \pTchjet~ range reported here at LHC beam energy, the inclusive jet sample is predominantly gluon-initiated, and thus the observed scaling behavior reflects the characteristic behavior of gluon jets. At higher jet energies, the quark-initiated jet fraction increases, which may lead to modifications in the scaling relation $\RL^{\text{peak}}\times \langle\pTchjet\rangle$. Testing this evolution and potential deviations from the observed scaling will be possible with the high-statistics data from ALICE Run 3.

At large \RL, a next-to-leading-logarithmic (NLL) pQCD calculation~\cite{Lee:2022ige} is shown in the orange curve.
The overall magnitude of the pQCD curve is normalized to data in the large-angle region where \\$\langle\pTchjet\rangle\RL~\gg~\Lambda_{\text{QCD}}$ ~\cite{Lee:2022ige}. The normalization range is \RL~between 12 \GeVc$/\langle\pTchjet\rangle$ (indicated by the dashed orange line in Fig.~\ref{fig:pT-shifted-compairison}) and 0.4 (jet radius), which is accessible in the measured EEC distribution of the 60--80 \GeVc~range. Inside this \RL~range, qualitative agreement is observed between the pQCD prediction and the measurement, except for the last point where \RL~approaches the jet radius. With higher $p_T$ jets, the \RL~range accessible for comparison with perturbative predictions extends to smaller values, as demonstrated in the CMS measurement at higher jet energies~\cite{CMS:2024mlf}.
At smaller \RL, the data deviate from perturbative scaling, which is expected as non-perturbative effects like soft radiation and hadronization become important. Recent theoretical studies show that power corrections for non-perturbative effects~\cite{Schindler:2023cww, Lee:2024esz} can be added to pQCD calculations, which extend the description of EEC measurements to smaller angles.

After the transition region where a broad peak structure is observed, the data follow a different \RL~dependence at very small \RL. This is can be described by a linear function in \RL, shown by the purple curve in Fig.~\ref{fig:pT-shifted-compairison}. The linear function is fit to the EEC distribution of 20--40 \GeVc~in \RL~range of 0.01, the lower limit of measured \RL~range, to 0.7 \GeVc/$\langle\pTchjet\rangle$, which is represented by the dashed purple line in Fig.~\ref{fig:pT-shifted-compairison}. Linear scaling is expected if the energy is uniformly distributed, where the correlation strength depends only on the area of the infinitesimal ring at distance \RL~and is thus proportional to $\RL d\RL$, which is often referred to as the non-perturbative or free hadron scaling region~\cite{Komiske:2022enw, CMS:2024mlf, STAR:2025jut}. This linear dependence, resulting from hadronization, breaks down towards larger angles because hadrons are produced from localized color fields, leading to a suppression of correlations between hadrons at large angular separations (\RL). In this region, parton splittings become the dominant contribution to the distribution.

The maxima of the transition peaks coincide at $\langle\pTchjet\rangle\RL = 2.43 \pm 0.14~\GeVc$. This observation implies a common energy scale where QCD radiations stop and hadronization effects take over, independent of the jet energy. Interestingly, the transition peak position is not affected by lowering the \pTtrack~selection to 0.15~\GeVc~for the measured \pTchjet~range (see Appendix~\ref{app:C} in Ref.~\cite{supplemental}). 
The shape of the transition region in Fig.~\ref{fig:pT-shifted-compairison} is also independent of jet energy, up to 80 \GeVc~- the highest \pT~accessible with our data sample.

\begin{figure}
    \centering
    \includegraphics[width=0.92\textwidth, height=0.5\textwidth]{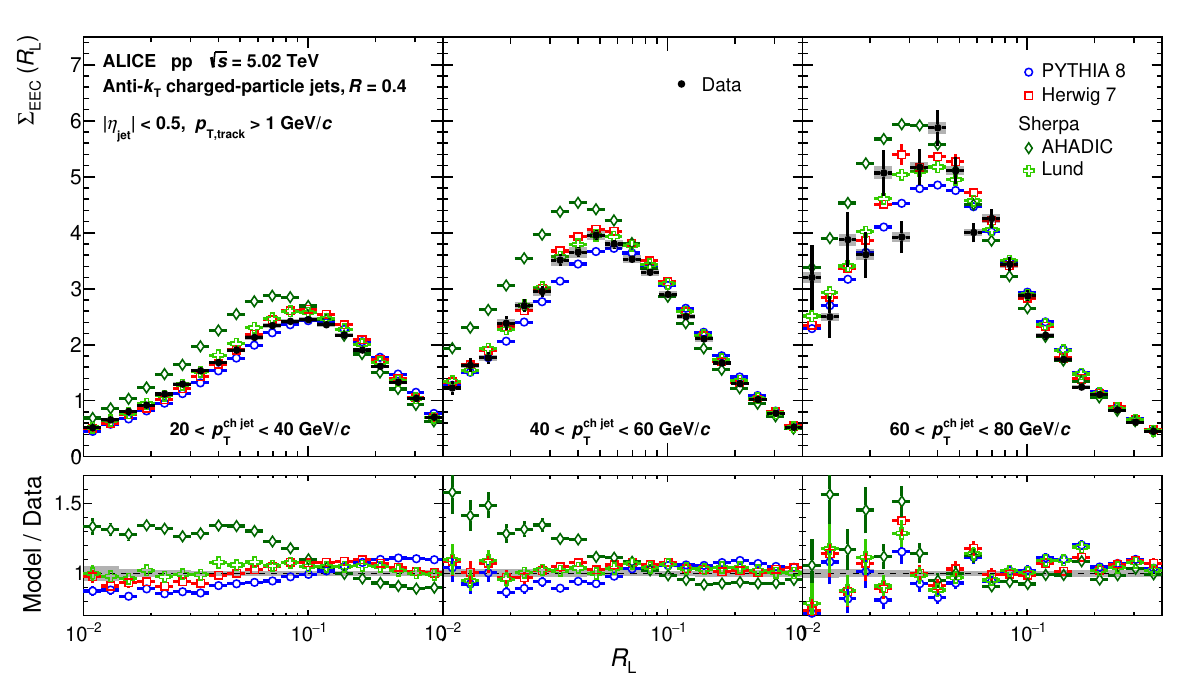}
    \caption{Top: Fully corrected \sigmaeec~as a function of \RL. Bottom: Ratio of \sigmaeec~from MC event generators to data. The gray band centered at 1 represents systematic uncertainty from data.}
    \label{fig:model-compairison}
\end{figure}

Fig.~\ref{fig:model-compairison} compares the measured EEC distributions and the MC models PYTHIA 8.235 (Monash tune)~\cite{Skands:2014pea}, Herwig 7.2.1 (default tune)~\cite{Bellm:2015jjp}, and Sherpa 2.2.15~\cite{Sherpa:2019gpd}. The default parton shower models are used for PYTHIA 8 (\pT-ordered), Herwig 7 (angular-ordered), and Sherpa (\pT-ordered). The MC models used here implement different hadronization mechanisms. PYTHIA 8 uses the Lund string hadronization model, which produces hadrons by breaking a color string into color-neutral hadrons.
Herwig 7 uses a cluster hadronization model, where locally color-connected partons are collected into clusters which then decay into hadrons. In Sherpa, the AHADIC tune uses a cluster hadronization model, while the Lund tune uses the Lund string hadronization. Both PYTHIA 8 and Herwig 7 describe the data within $20\%$ across the measured \RL~range. The transition region peaks at a slightly larger angle in PYTHIA 8 compared to Herwig 7. Herwig 7 better describes the transition peak position, but overestimates the maximum value of the
transition peak. For Sherpa, the Lund tune agrees with the data quite well, while the AHADIC tune shows up to a 50$\%$ deviation in the small-angle region. The agreement with data improves for MC models as the jet \pT~increases. When comparing the EEC distributions from Sherpa with different hadronization models to data, it is worth noting that the transition region from AHADIC (cluster hadronization) is at a smaller angle compared to the Lund tune (string hadronization). Such a shift of the transition region is also observed when comparing Herwig 7 and PYTHIA 8, where Herwig 7 shows a transition region at smaller angles compared to PYTHIA 8. The impact of the different hadronization parameters in PYTHIA 8 on the transition region of EEC distributions can be found in Appendix~\ref{app:D}. More detailed comparisons with MC models could constrain the parton shower, hadronization, and hadronic interactions in the models.

\section{Conclusions}
\label{sec:conclusions}

We report the EEC distributions for charged-particle jets produced in pp collisions measured with the ALICE detector. This marks the first energy correlator measurements that are fully corrected for detector effects at the LHC for low-\pT\ jets. The data are in good agreement with a pQCD calculation in the large angle region. EEC at small \RL~can be described by a linear function of \RL, consistent with a spatially uniform population of particles produced from hadronization. The transition region is established at the crossover of the two mechanisms, with perturbative contributions diminishing toward smaller \RL~and hadronization contributions diminishing toward larger~\RL. The transition region shows a clear jet energy dependence, peaking at a smaller angular scale for jets with higher \pT. The re-normalized EEC distributions as a function of $\langle\pTchjet\rangle\RL$ reveal that this quantity governs the shape of the EEC distribution, directly relating to the energy scale (virtuality) of jets. A common transition position occurs at $\RL \sim(2.4~\GeVc)/\langle\pTchjet\rangle$. 

MC models describe the data within $20\%$ across the measured \RL~range, except for Sherpa with AHADIC tune. The discrepancy between models and data decreases for jets with higher energy. As the hadronization process is a non-perturbative phenomenon, theory predictions based on first principles are challenging, and developments in lattice QCD and quantum computation may offer needed improvements~\cite{deJong:2021wsd}. In the future, measurements of higher-order energy correlators and comparison of EEC measurements from different collision systems can provide further insight on the confinement process~\cite{Tkachov:1995kk, Komiske:2022enw, Chen:2020vvp}.
The study presented in this letter provides a baseline for using EEC to study cold and hot nuclear medium effects on jet formation and propagation.

\newenvironment{acknowledgement}{\relax}{\relax}
\begin{acknowledgement}
\section*{Acknowledgements}
The ALICE Collaboration gratefully acknowledges Kyle Lee for providing theoretical predictions. 
%
The ALICE Collaboration would like to thank all its engineers and technicians for their invaluable contributions to the construction of the experiment and the CERN accelerator teams for the outstanding performance of the LHC complex.
The ALICE Collaboration gratefully acknowledges the resources and support provided by all Grid centres and the Worldwide LHC Computing Grid (WLCG) collaboration.
The ALICE Collaboration acknowledges the following funding agencies for their support in building and running the ALICE detector:
A. I. Alikhanyan National Science Laboratory (Yerevan Physics Institute) Foundation (ANSL), State Committee of Science and World Federation of Scientists (WFS), Armenia;
Austrian Academy of Sciences, Austrian Science Fund (FWF): [M 2467-N36] and Nationalstiftung f\"{u}r Forschung, Technologie und Entwicklung, Austria;
Ministry of Communications and High Technologies, National Nuclear Research Center, Azerbaijan;
Conselho Nacional de Desenvolvimento Cient\'{\i}fico e Tecnol\'{o}gico (CNPq), Financiadora de Estudos e Projetos (Finep), Funda\c{c}\~{a}o de Amparo \`{a} Pesquisa do Estado de S\~{a}o Paulo (FAPESP) and Universidade Federal do Rio Grande do Sul (UFRGS), Brazil;
Bulgarian Ministry of Education and Science, within the National Roadmap for Research Infrastructures 2020-2027 (object CERN), Bulgaria;
Ministry of Education of China (MOEC) , Ministry of Science \& Technology of China (MSTC) and National Natural Science Foundation of China (NSFC), China;
Ministry of Science and Education and Croatian Science Foundation, Croatia;
Centro de Aplicaciones Tecnol\'{o}gicas y Desarrollo Nuclear (CEADEN), Cubaenerg\'{\i}a, Cuba;
Ministry of Education, Youth and Sports of the Czech Republic, Czech Republic;
The Danish Council for Independent Research | Natural Sciences, the VILLUM FONDEN and Danish National Research Foundation (DNRF), Denmark;
Helsinki Institute of Physics (HIP), Finland;
Commissariat \`{a} l'Energie Atomique (CEA) and Institut National de Physique Nucl\'{e}aire et de Physique des Particules (IN2P3) and Centre National de la Recherche Scientifique (CNRS), France;
Bundesministerium f\"{u}r Forschung, Technologie und Raumfahrt (BMFTR) and GSI Helmholtzzentrum f\"{u}r Schwerionenforschung GmbH, Germany;
General Secretariat for Research and Technology, Ministry of Education, Research and Religions, Greece;
National Research, Development and Innovation Office, Hungary;
Department of Atomic Energy Government of India (DAE), Department of Science and Technology, Government of India (DST), University Grants Commission, Government of India (UGC) and Council of Scientific and Industrial Research (CSIR), India;
National Research and Innovation Agency - BRIN, Indonesia;
Istituto Nazionale di Fisica Nucleare (INFN), Italy;
Japanese Ministry of Education, Culture, Sports, Science and Technology (MEXT) and Japan Society for the Promotion of Science (JSPS) KAKENHI, Japan;
Consejo Nacional de Ciencia (CONACYT) y Tecnolog\'{i}a, through Fondo de Cooperaci\'{o}n Internacional en Ciencia y Tecnolog\'{i}a (FONCICYT) and Direcci\'{o}n General de Asuntos del Personal Academico (DGAPA), Mexico;
Nederlandse Organisatie voor Wetenschappelijk Onderzoek (NWO), Netherlands;
The Research Council of Norway, Norway;
Pontificia Universidad Cat\'{o}lica del Per\'{u}, Peru;
Ministry of Science and Higher Education, National Science Centre and WUT ID-UB, Poland;
Korea Institute of Science and Technology Information and National Research Foundation of Korea (NRF), Republic of Korea;
Ministry of Education and Scientific Research, Institute of Atomic Physics, Ministry of Research and Innovation and Institute of Atomic Physics and Universitatea Nationala de Stiinta si Tehnologie Politehnica Bucuresti, Romania;
Ministry of Education, Science, Research and Sport of the Slovak Republic, Slovakia;
National Research Foundation of South Africa, South Africa;
Swedish Research Council (VR) and Knut \& Alice Wallenberg Foundation (KAW), Sweden;
European Organization for Nuclear Research, Switzerland;
Suranaree University of Technology (SUT), National Science and Technology Development Agency (NSTDA) and National Science, Research and Innovation Fund (NSRF via PMU-B B05F650021), Thailand;
Turkish Energy, Nuclear and Mineral Research Agency (TENMAK), Turkey;
National Academy of  Sciences of Ukraine, Ukraine;
Science and Technology Facilities Council (STFC), United Kingdom;
National Science Foundation of the United States of America (NSF) and United States Department of Energy, Office of Nuclear Physics (DOE NP), United States of America.
In addition, individual groups or members have received support from:
Czech Science Foundation (grant no. 23-07499S), Czech Republic;
FORTE project, reg.\ no.\ CZ.02.01.01/00/22\_008/0004632, Czech Republic, co-funded by the European Union, Czech Republic;
European Research Council (grant no. 950692), European Union;
ICSC - Centro Nazionale di Ricerca in High Performance Computing, Big Data and Quantum Computing, European Union - NextGenerationEU;
Academy of Finland (Center of Excellence in Quark Matter) (grant nos. 346327, 346328), Finland. 
\end{acknowledgement}

\bibliographystyle{utphys}   
\bibliography{main}

\newpage
\appendix
\section{Additional analysis information}
\label{app:new}

\begin{figure}[!htb]
\centering{}
\includegraphics[scale=0.39]{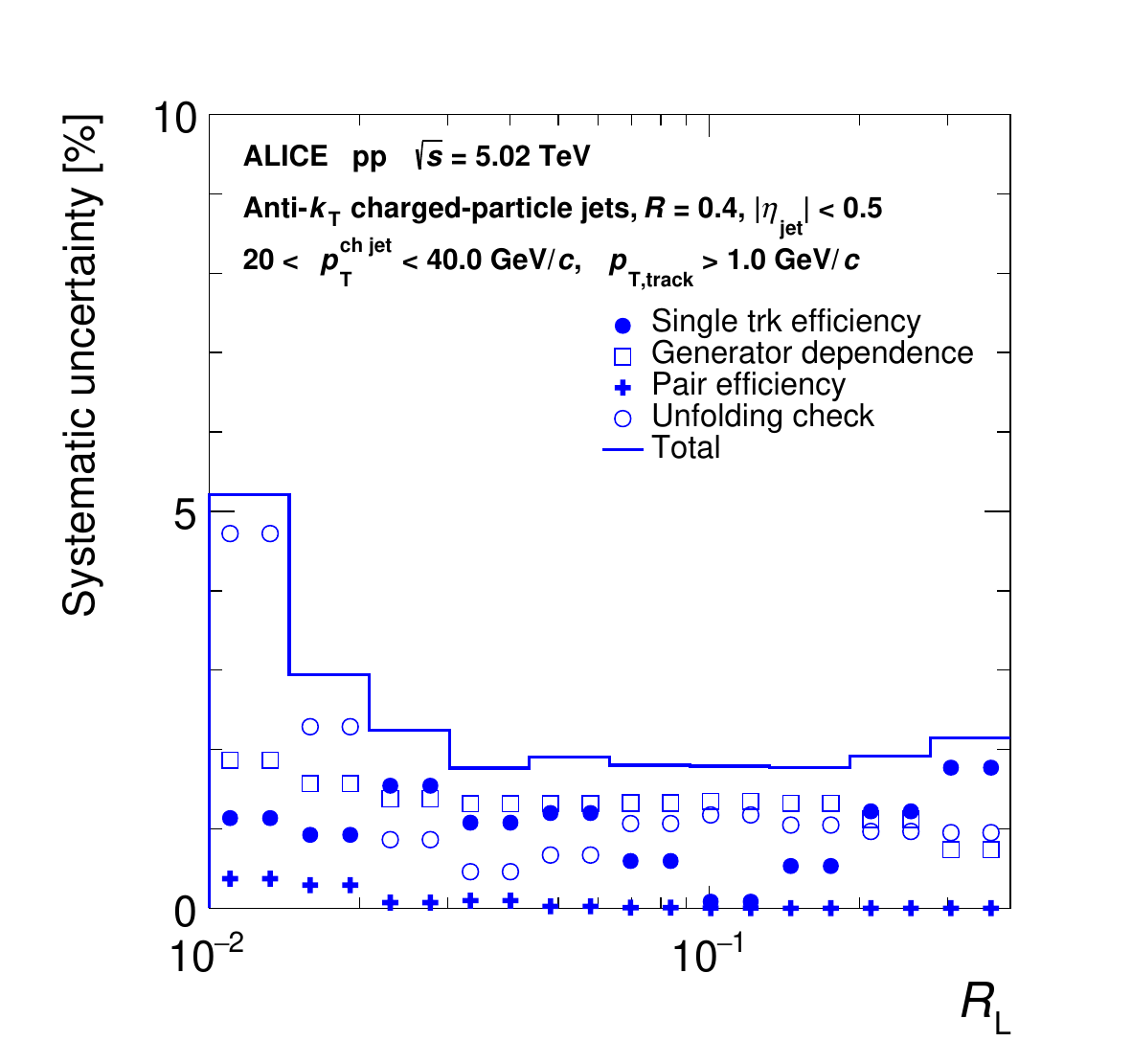}
\includegraphics[scale=0.39]{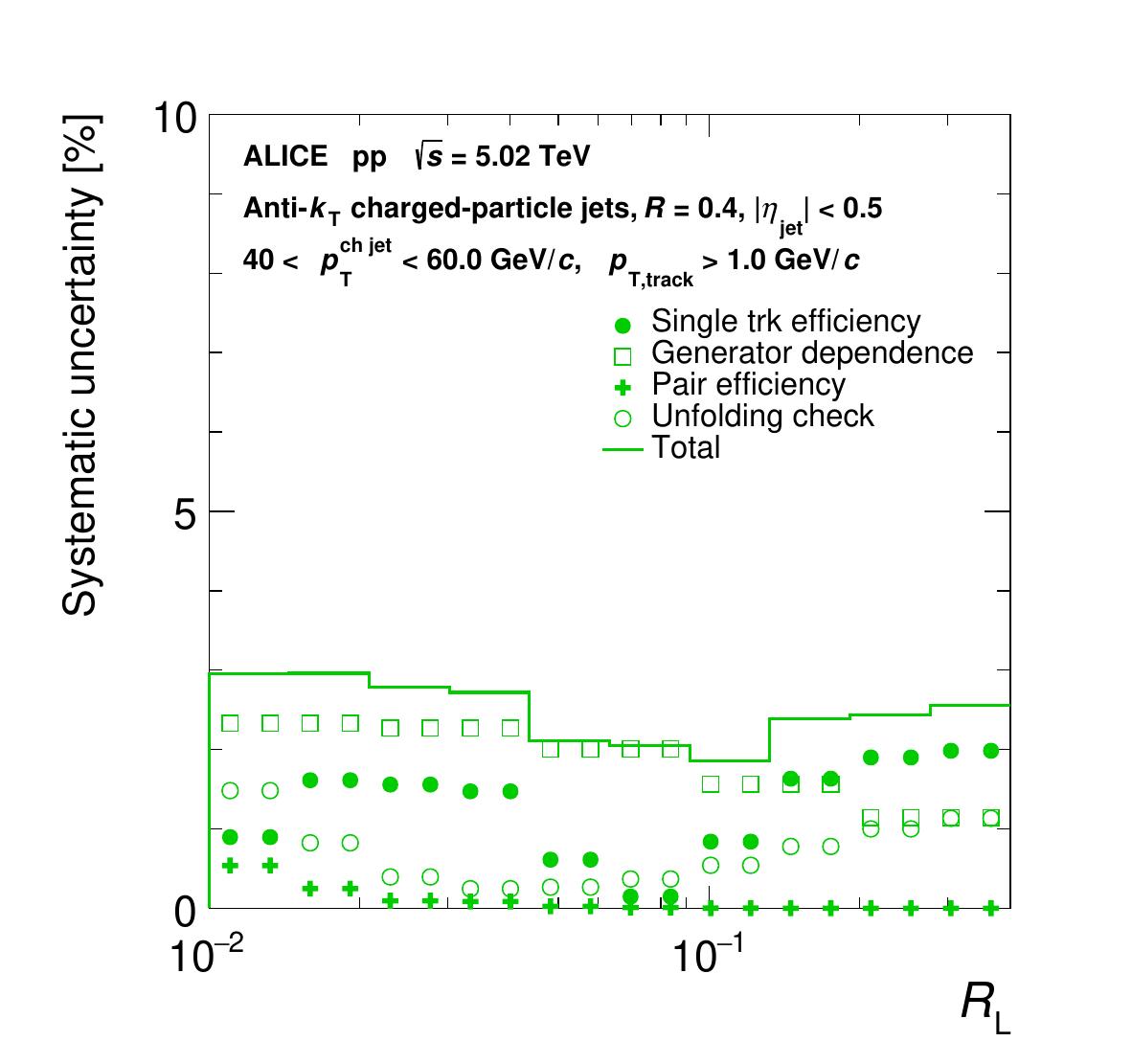}
\includegraphics[scale=0.39]{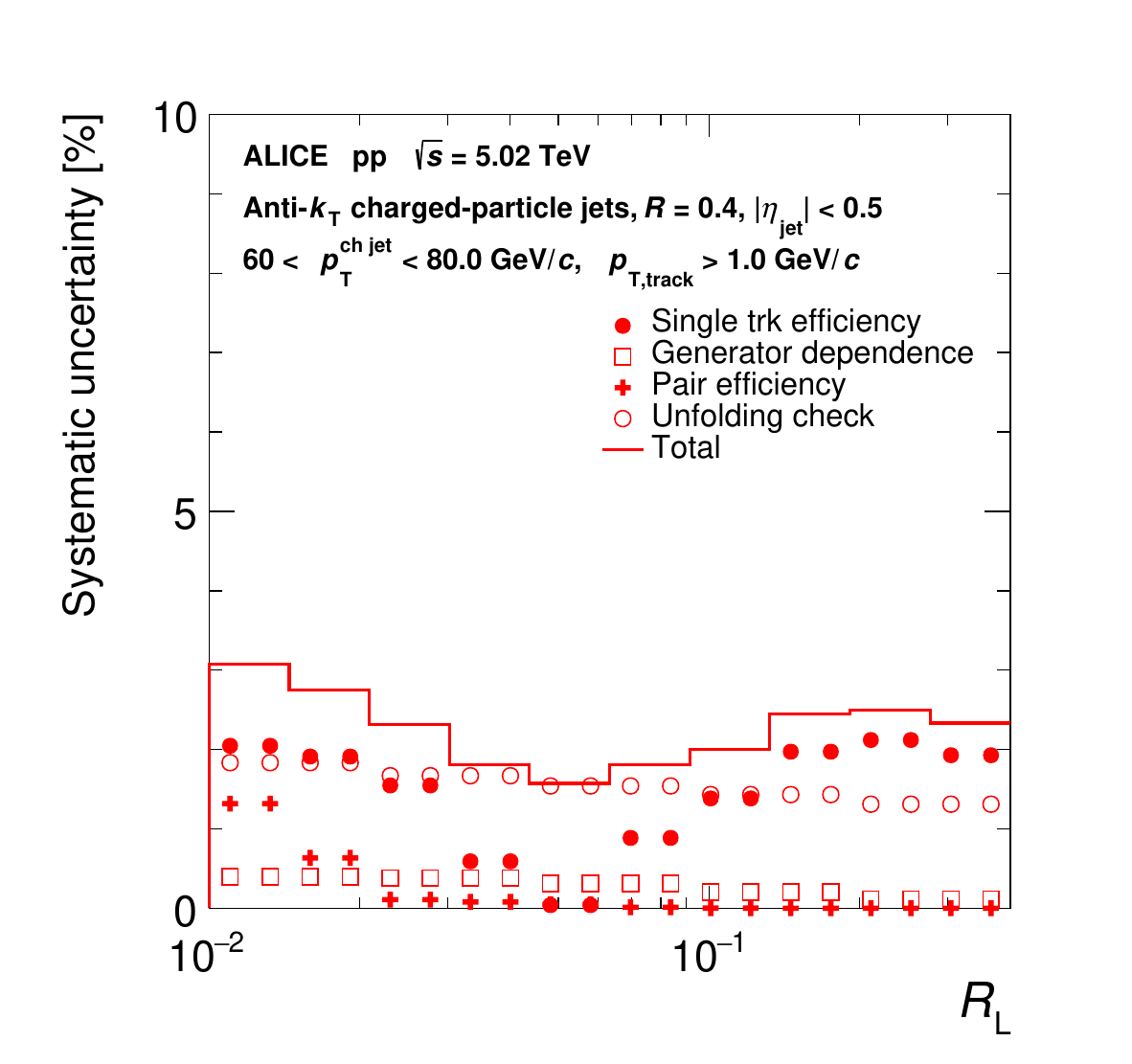}
\caption{Systematic uncertainties for fully corrected $\Sigma_{\text{EEC}}$ as a function of $R_L$ in the \pTchjet{} intervals 20--40, 40--60, and 60--80~\GeVc}
\label{fig:sys-err}
\end{figure}

Fig.~\ref{fig:sys-err} shows a detailed breakdown of the systematic uncertainties from individual sources.

The unfolding is performed in three dimensions for EEC pairs inside jets similar to the CMS measurement~\cite{CMS:2024mlf}, \RL, \pTchjet{}, and energy weight. The same simulation setup used to derive the bin-by-bin correction factors is used to construct the response matrices. The energy weights are divided into 20 equal intervals on a logarithmic scale, ranging from $10^{-5}$ to 1. The unfolded energy weight is then summed to calculate the $\Sigma_{\text{EEC}}$ using the following equation:
\begin{equation}
    \Sigma_{\rm EEC} (R_{\rm L}, \pTchjet{}) = \sum_{w_{\rm pair}~\rm bins}w_{\rm pair} \times N_{\rm pair}(w_{\rm pair}, R_{\rm L}, \pTchjet{})
    \label{eq:sum_weights}
\end{equation}
where $w_{\rm pair}$ represents a certain energy weight bin, and $N_{\rm pair}(w_{\rm pair}, R_{\rm L}, \pTchjet{})$ is the number of pairs in the corresponding energy weight bin. 

The finite number of bins used for energy weight in the unfolding method results in a slightly higher $\Sigma_{\text{EEC}}$ than the truth value obtained from directly summing over weights of all EEC pairs, which is $\approx1.2\%$ for our binning choice. To account for such effect with our binning choice, a multiplicative correction factors are applied on the $\Sigma_{\text{EEC}}$ calculated from Eq.~\ref{eq:sum_weights}.

\section{Characterization of the transition region} \label{app:B}

To quantitatively characterize the transition region, we employ two complementary fitting approaches to extract the transition peak position and peak height, as detailed below.

\textbf{Method 1: Local log-normal fits on the EEC distributions}

\begin{figure}[!htb]
\centering{}
\includegraphics[scale=0.39]{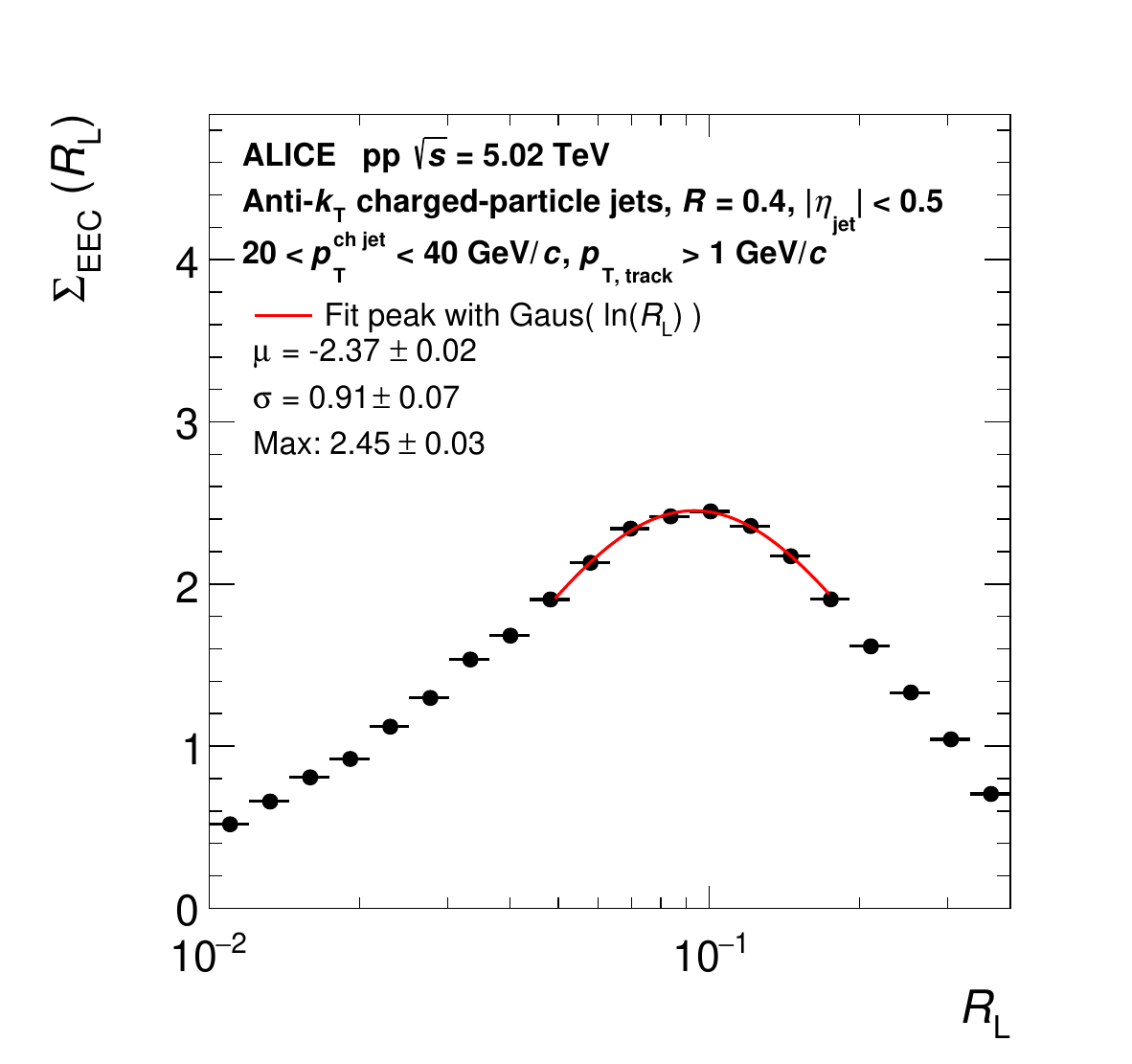}
\includegraphics[scale=0.39]{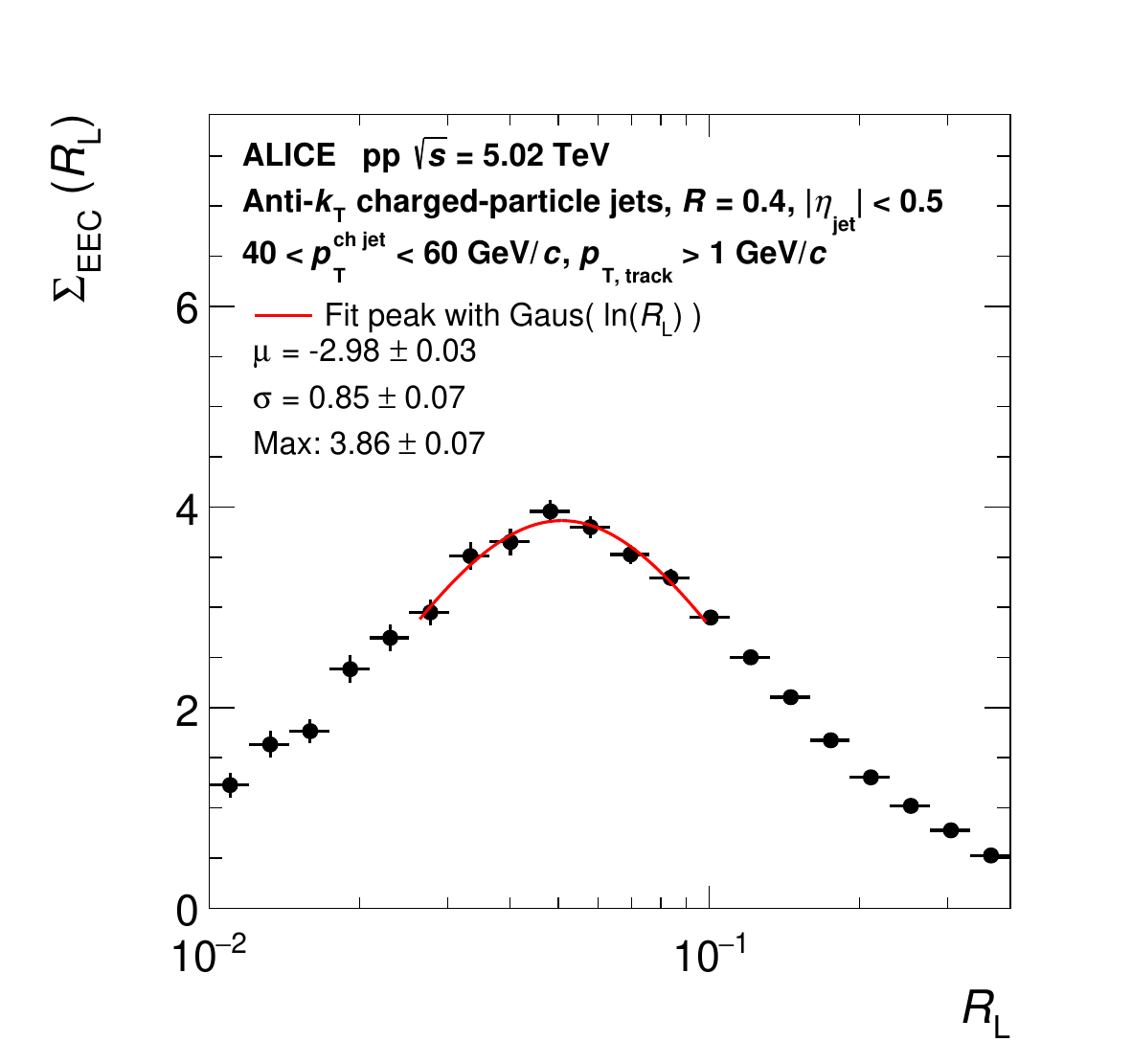}
\includegraphics[scale=0.39]{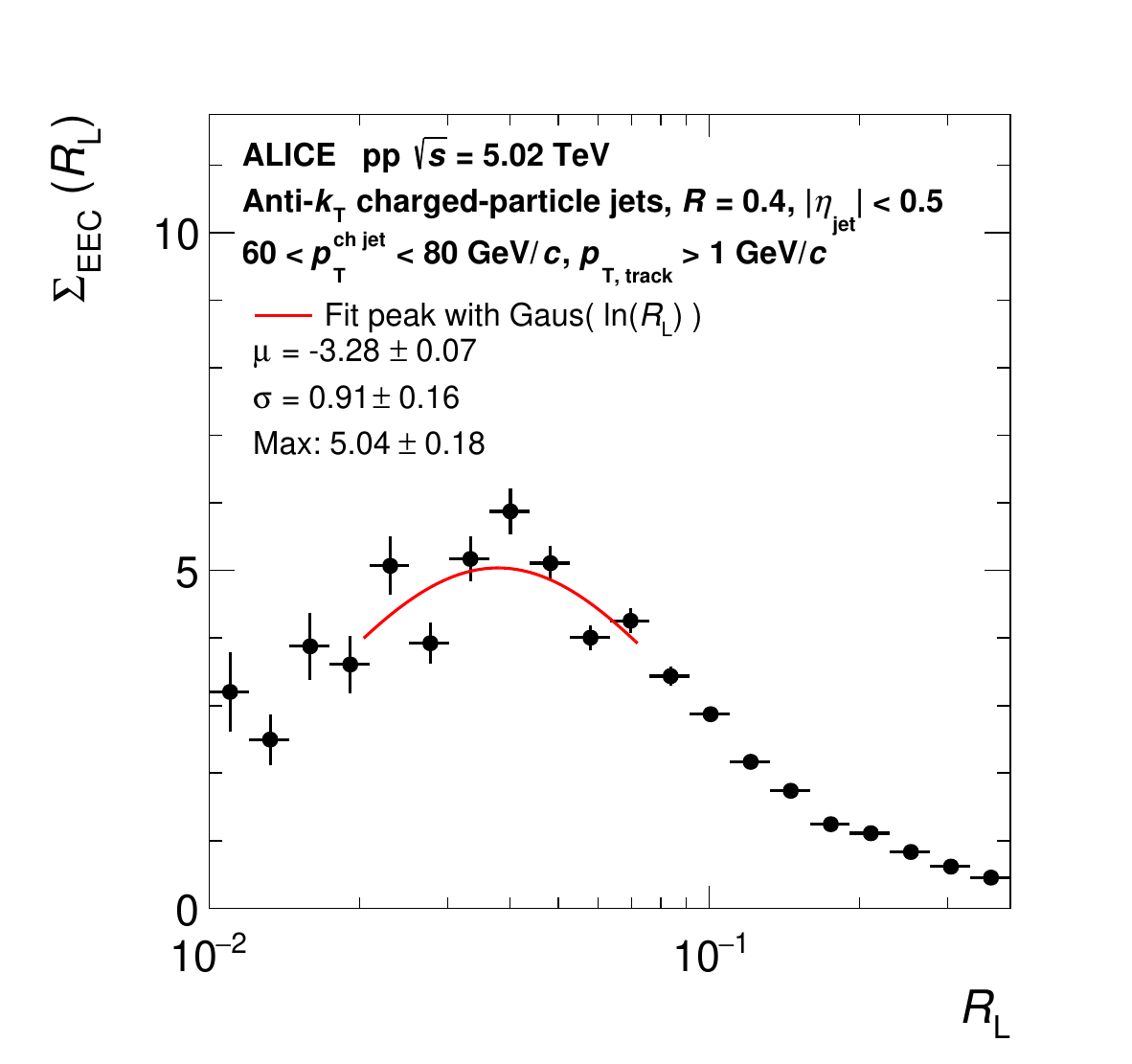}
\caption{Extraction of the peak position and height of the EEC distributions in different jet \pT~ranges using local log-normal fits. The fit function and fitting range are shown by the red curves.}
\label{fig:fitEEC}
\end{figure}

We perform local fits to the EEC distributions in each jet \pT~range separately using a log-normal function:
\begin{equation}
\text{Gaus}(\ln(\RL)) = C\cdot\exp\left(-\frac{(\ln \RL - \mu)^2}{2\sigma^2}\right)
\end{equation}
where $C$ is a normalization constant. The fit range is restricted to $\mu\pm0.7\sigma$, the region where the function provides a reasonable description of the transition peak. The fit results are shown in Fig.~\ref{fig:fitEEC}. The peak position $\RL^{\text{peak}}$ and its uncertainty are extracted via $\RL^{\text{peak}} = \exp(\mu)$.

\begin{figure}[!htb]
\centering{}
\includegraphics[scale=0.38]{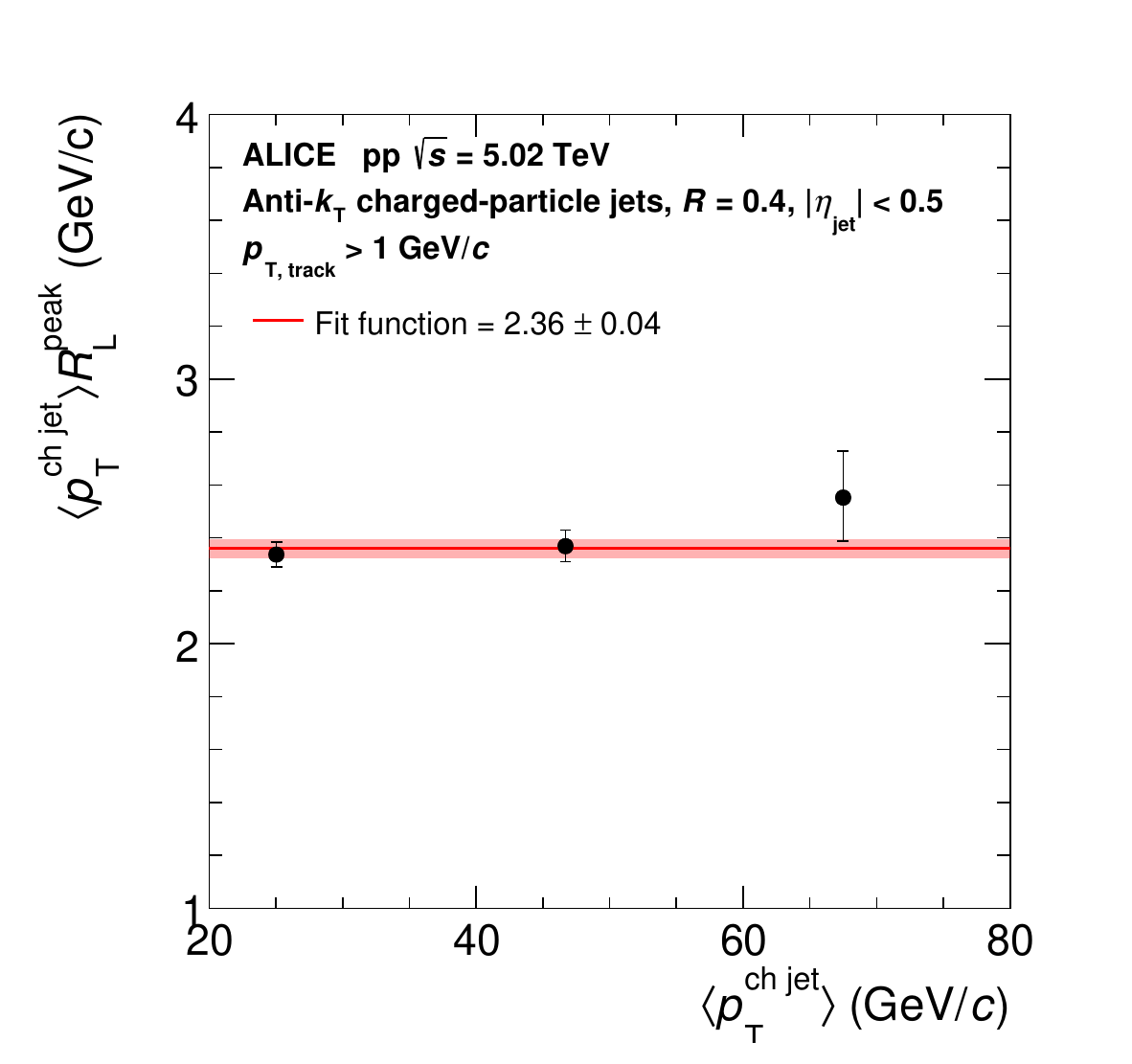}
\includegraphics[scale=0.38]{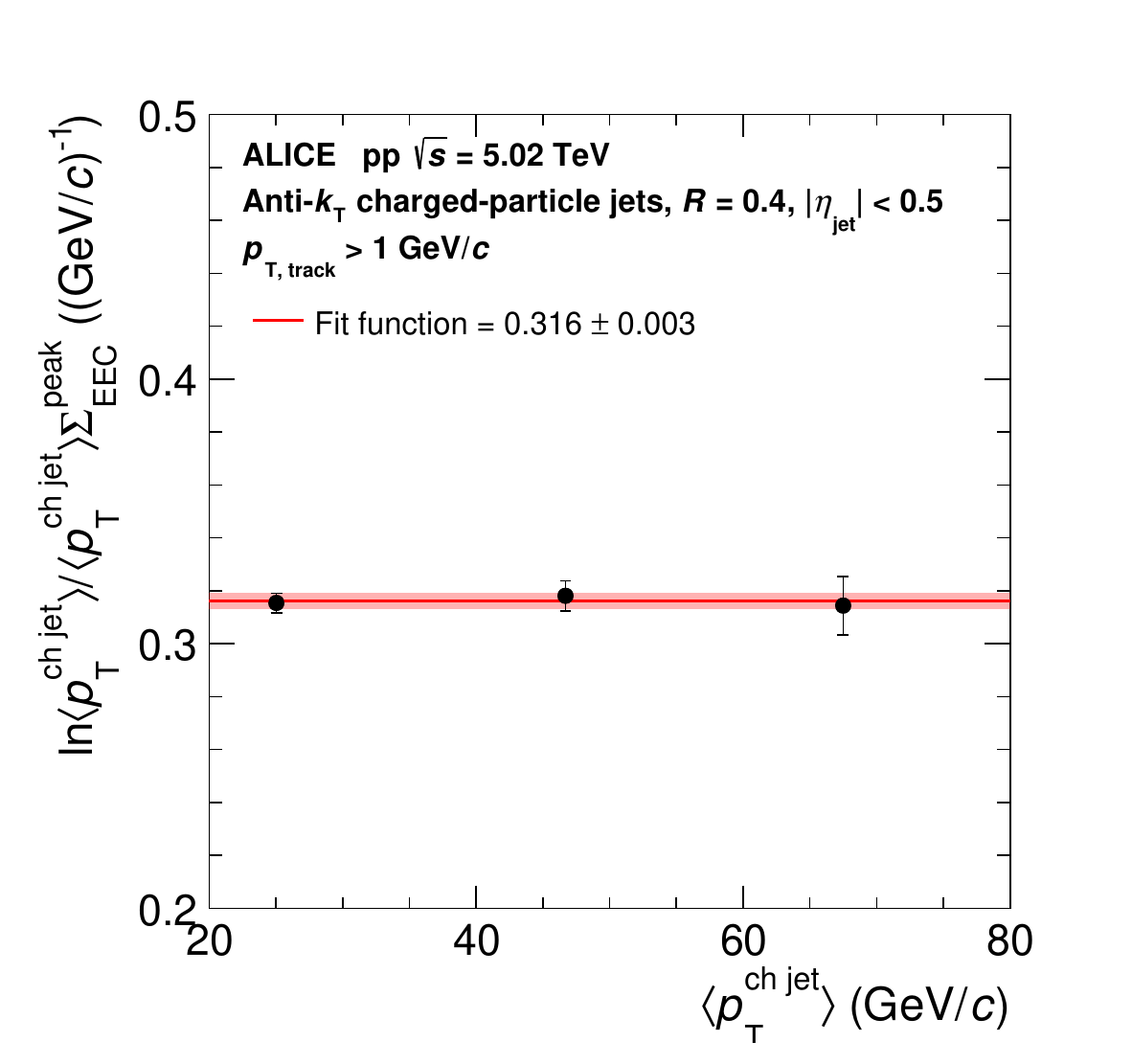}
\caption{Left: jet \pT~dependence of the transition peak position $\RL^{\text{peak}}$. Right: jet \pT~dependence of the transition peak height $\sigmaeec^{\text{peak}}$. The notation $\ln\langle\pTchjet\rangle$ in the y-axis represents $\ln(\langle\pTchjet\rangle/(\GeVc))$ as explained in footnote~\ref{fn:lnpt}.}
\label{fig:scaled_peak_pos_height}
\end{figure}

Using the extracted values from Fig.~\ref{fig:fitEEC}, we examine the jet \pT~dependence of the peak position and peak height in Fig.~\ref{fig:scaled_peak_pos_height}. The mean values $\langle\pTchjet\rangle$ for the three jet \pTchjet~intervals (20--40, 40--60, 60--80~\GeVc) are 25.1, 46.7, and 67.5~\GeVc, respectively. The notation $\ln\langle\pTchjet\rangle$ represents $\ln(\langle\pTchjet\rangle/(\GeVc))$ as explained in footnote~\ref{fn:lnpt}. Using this method, we obtain $\RL^{\text{peak}}\times \langle\pTchjet\rangle \approx 2.36 \pm 0.04~\GeVc$ and $\sigmaeec^{\text{peak}}\times \ln\langle\pTchjet\rangle/\langle\pTchjet\rangle \approx 0.316 \pm 0.003~(\GeVc)^{-1}$.

\textbf{Method 2: Global fit to scaled EEC distributions}

As an alternative approach, we employ the following functional form:
\begin{equation}
    f(x) = C\frac{x}{(x^2+T^{3/2})}
\end{equation}
where $C$ and $T$ are fit parameters and the maximum occurs at $x = \sqrt{T/2}$. This function exhibits linear dependence at small $x$, $x^{-2}$ behavior at large $x$, and a smooth transition between these limits, making it well-suited to describe the full shape of the transition region~\cite{STAR:2025jut}.

\begin{figure}[!htb]
\centering{}
\includegraphics[scale=0.45]{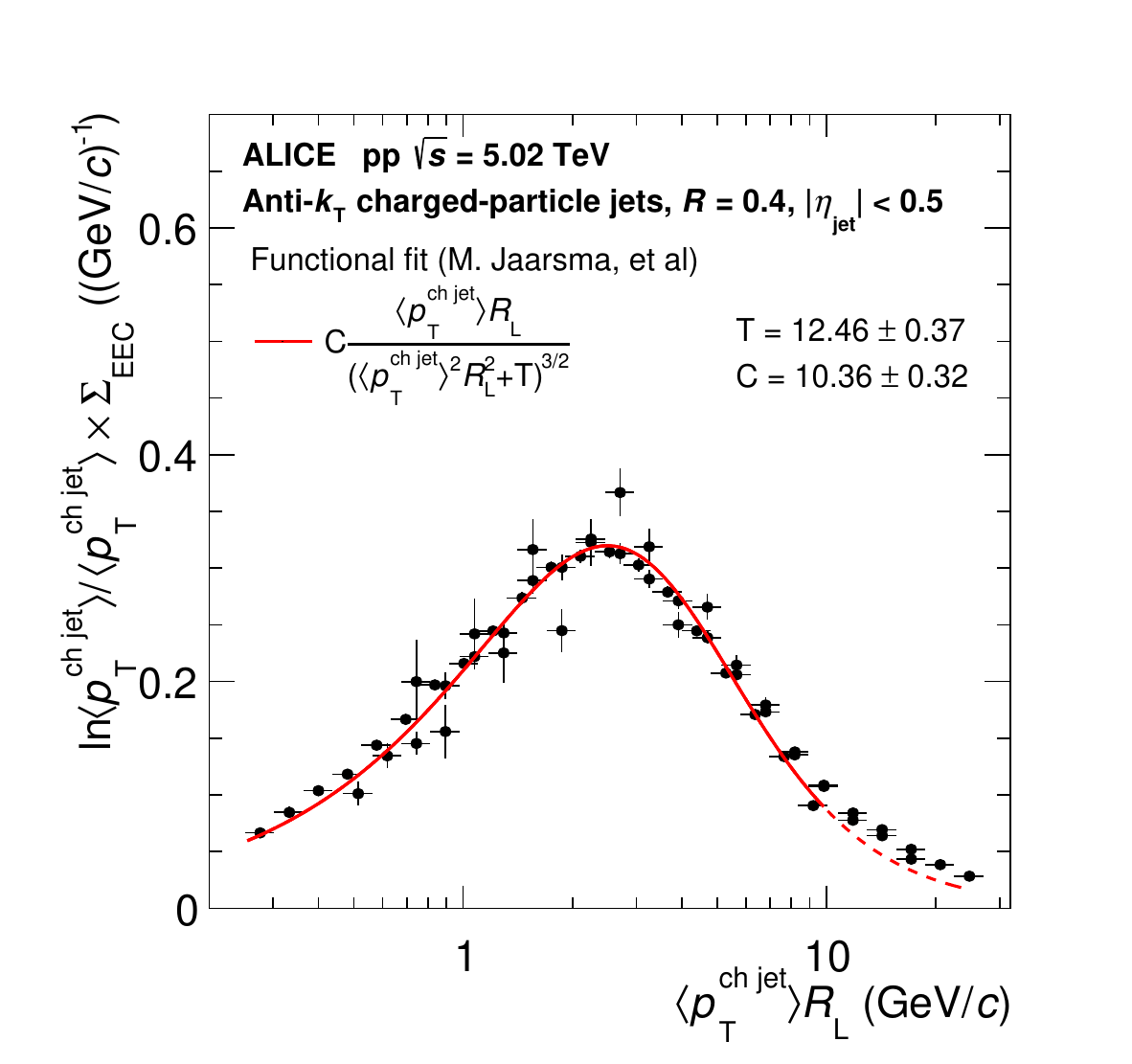}
\caption{Normalized $\Sigma_{\text{EEC}}$ as a function of $\langle\pTchjet\rangle\RL$ for \pTchjet~in 20--40, 40--60, and 60--80 \GeVc. Statistical and systematic uncertainties are combined in quadrature. The fit function is shown by the red curve, with the solid portion indicating the fitting range.}
\label{fig:fitEEC_combined}
\end{figure}

We fit this function to the combined \pT~scaled EEC distribution from all three jet \pT~ranges, which is shown in Fig.~\ref{fig:fitEEC_combined}. The fitting range is restricted to exclude the perturbative region at large $\langle\pTchjet\rangle\RL$, where this functional form is not expected to apply. From this global fit, we extract $\RL^{\text{peak}}\times \langle\pTchjet\rangle = \sqrt{T/2} \approx 2.50 \pm 0.04~\GeVc$ and $\sigmaeec^{\text{peak}}\times \ln\langle\pTchjet\rangle/\langle\pTchjet\rangle \approx 0.320 \pm 0.003~(\GeVc)^{-1}$.

\textbf{Final results}

To obtain robust estimates of the peak position and height, we average the results from both methods and assign the maximum difference between them as the uncertainty:
\begin{equation}
    \RL^{\text{peak}}\times \langle\pTchjet\rangle \approx 2.43 \pm 0.14~\GeVc
\label{eq:peak_pos}
\end{equation}
\begin{equation}
    \sigmaeec^{\text{peak}}\times \ln\langle\pTchjet\rangle/\langle\pTchjet\rangle \approx 0.314 \pm 0.004~(\GeVc)^{-1}
\label{eq:peak_height}
\end{equation}

The jet \pT~dependence of the transition region described in Eq.\ref{eq:peak_pos} and~\ref{eq:peak_height} can also be viewed in Fig.~\ref{fig:pT-compairison}.

\section{Results with different track threshold selections} \label{app:C}

\begin{figure}[!htb]
\centering{}
\includegraphics[width=0.47\textwidth]{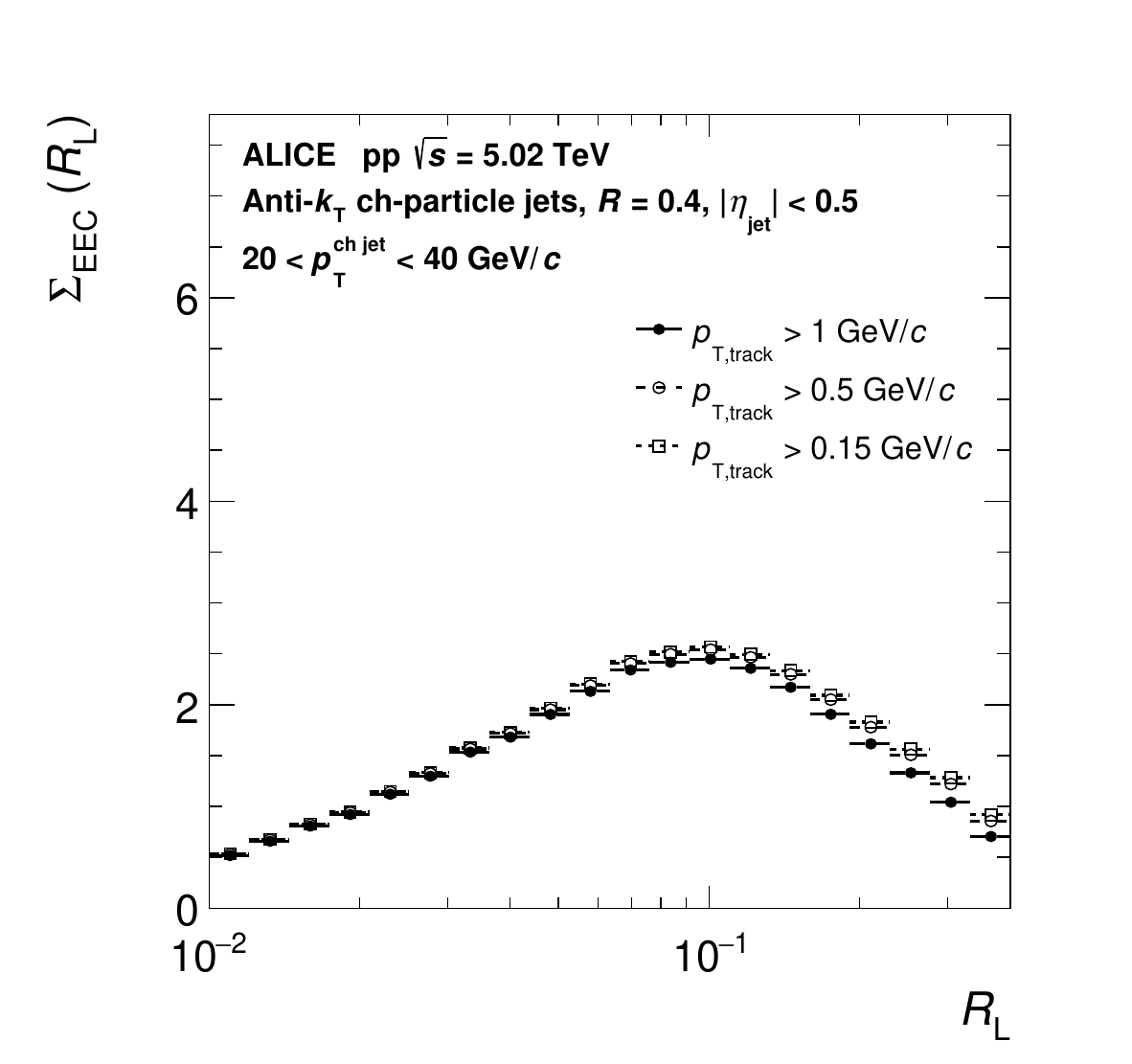}
\includegraphics[width=0.47\textwidth]{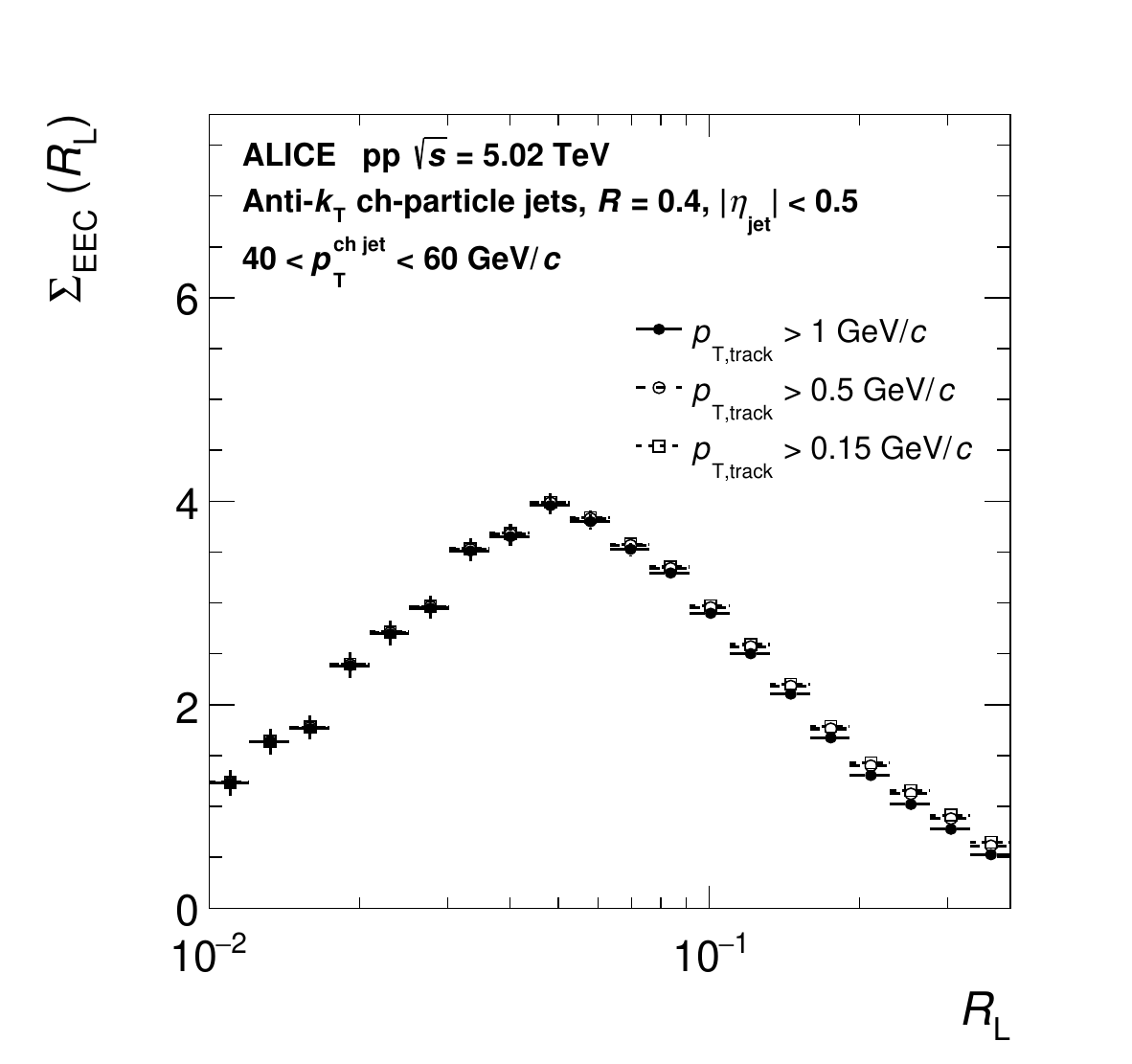}
\includegraphics[width=0.47\textwidth]{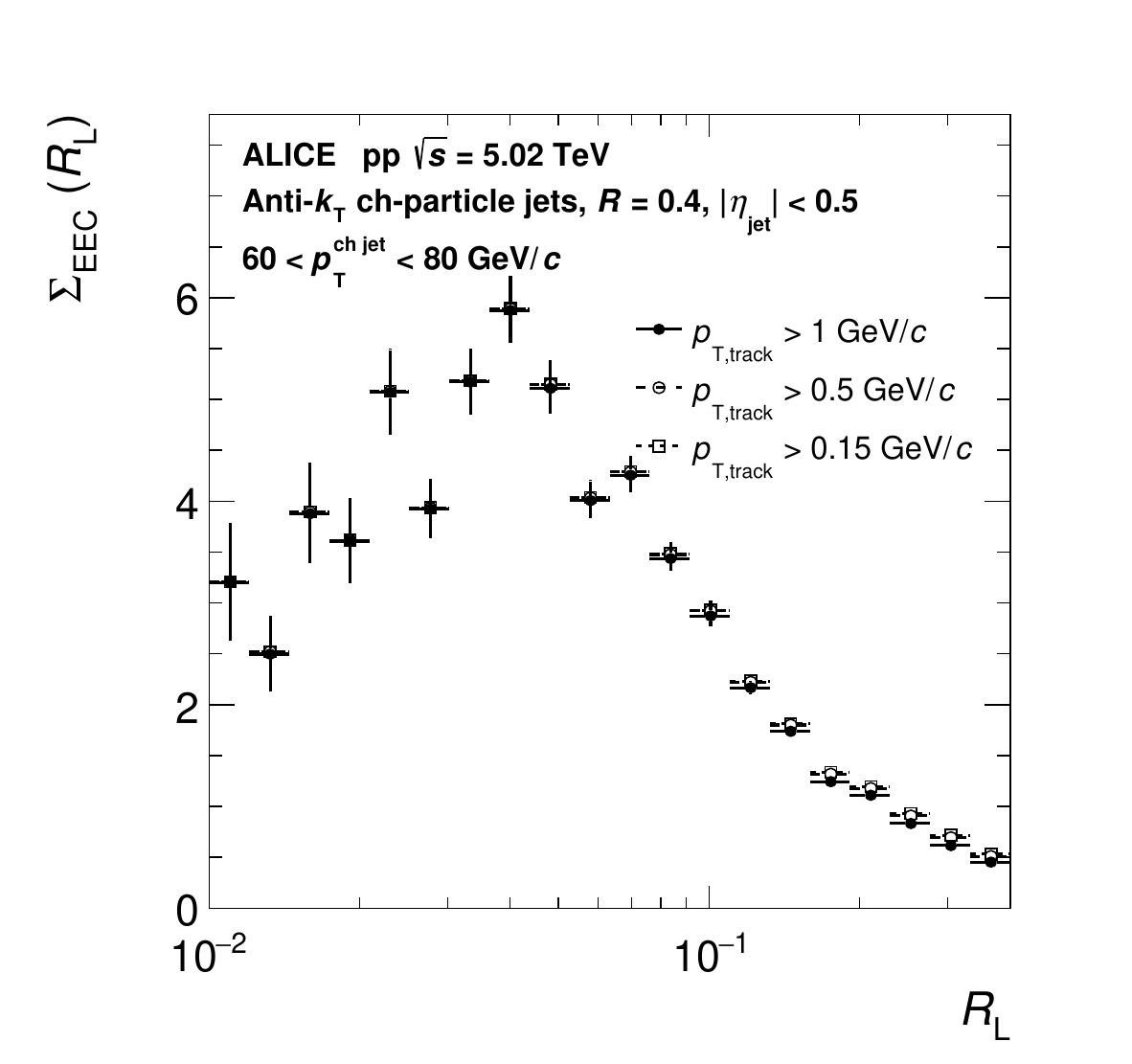}
\caption{Comparison \sigmaeec~as a function of \RL~with 0.15~\GeVc, 0.5 \GeVc, and 1~\GeVc~track \pT~threshold cut in the \pTchjet{} intervals 20--40, 40--60, and 60--80~\GeVc. Systematic uncertainties are not shown here.}
\label{fig:diff_thrd_corrected_EEC}
\end{figure}

The reported results in the main text use charged-particle tracks with $\pTtrack >$~1 \GeVc~threshold cut when constructing the EECs. We also carried out the measurement with lower threshold cuts, $\pTtrack > 0.15$~\GeVc~and $\pTtrack >$~0.5~\GeVc. The comparison of the EEC distributions with different track threshold selections is shown in Fig.~\ref{fig:diff_thrd_corrected_EEC}. Different track threshold selections lead to a small difference in the EEC distributions at large \RL~and the difference becomes negligible for jets with higher \pT. The transition peak positions are not affected by the different \pTtrack~selections used.

\section{PYTHIA 8 studies} \label{app:D}

To obtain more intuition about the transition region in the EEC distributions and the hadronization process, we studied with PYTHIA 8 (8.235 version) Monash tune how different parameters impact the EECs. The parameters checked are the following:
\begin{itemize}
\item Parameter related to the hadron fragmentation: StringPT:sigma, which corresponds to the average transverse momentum from string breaking.
\item Parameter related to the parton shower: TimeShower:pTmin, which corresponds to the parton shower cut-off \pT~for QCD emissions.
\end{itemize}

\begin{figure}[!htb]
\centering{}
\includegraphics[scale=0.39]{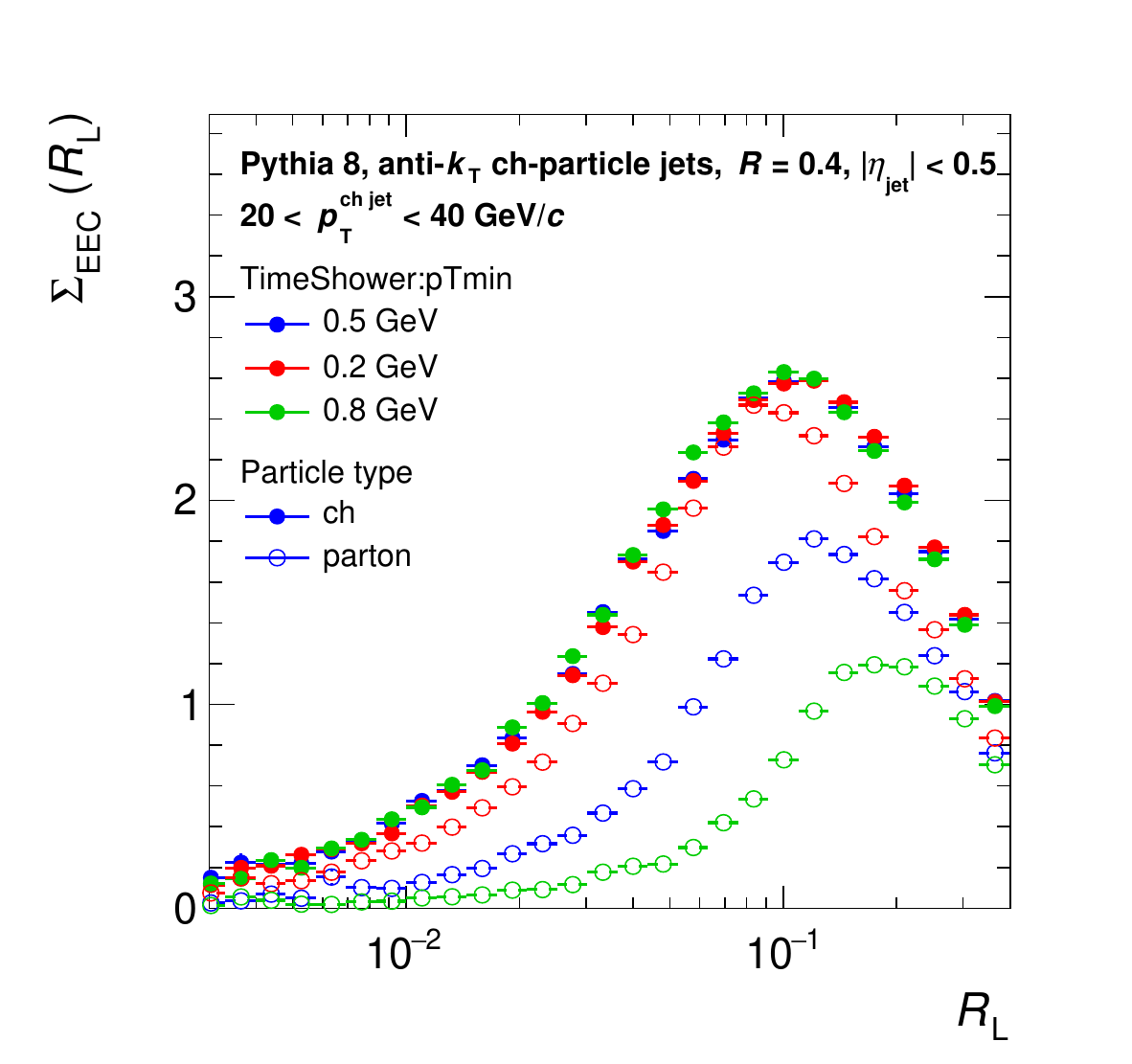}
\includegraphics[scale=0.39]{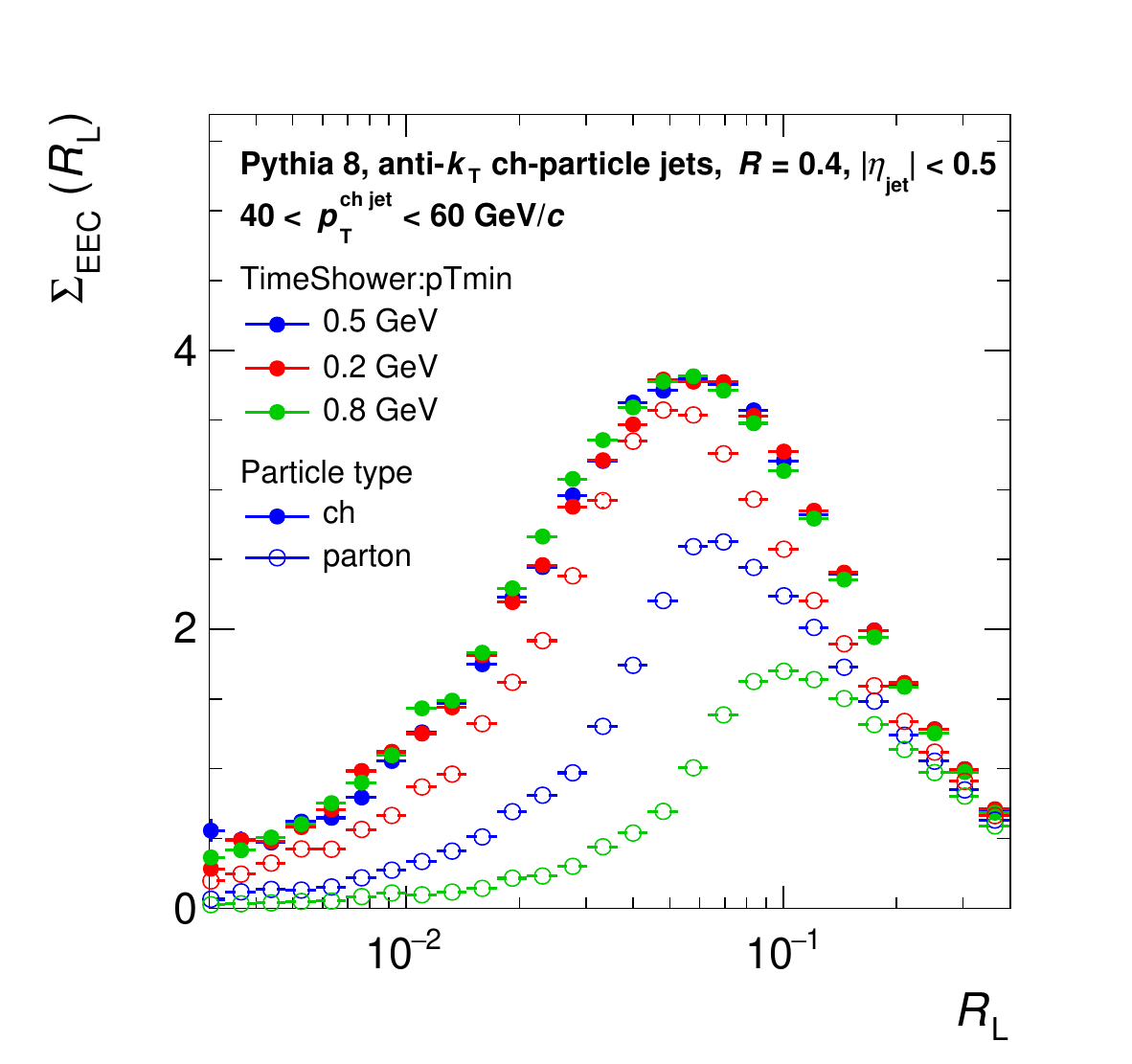}
\includegraphics[scale=0.39]{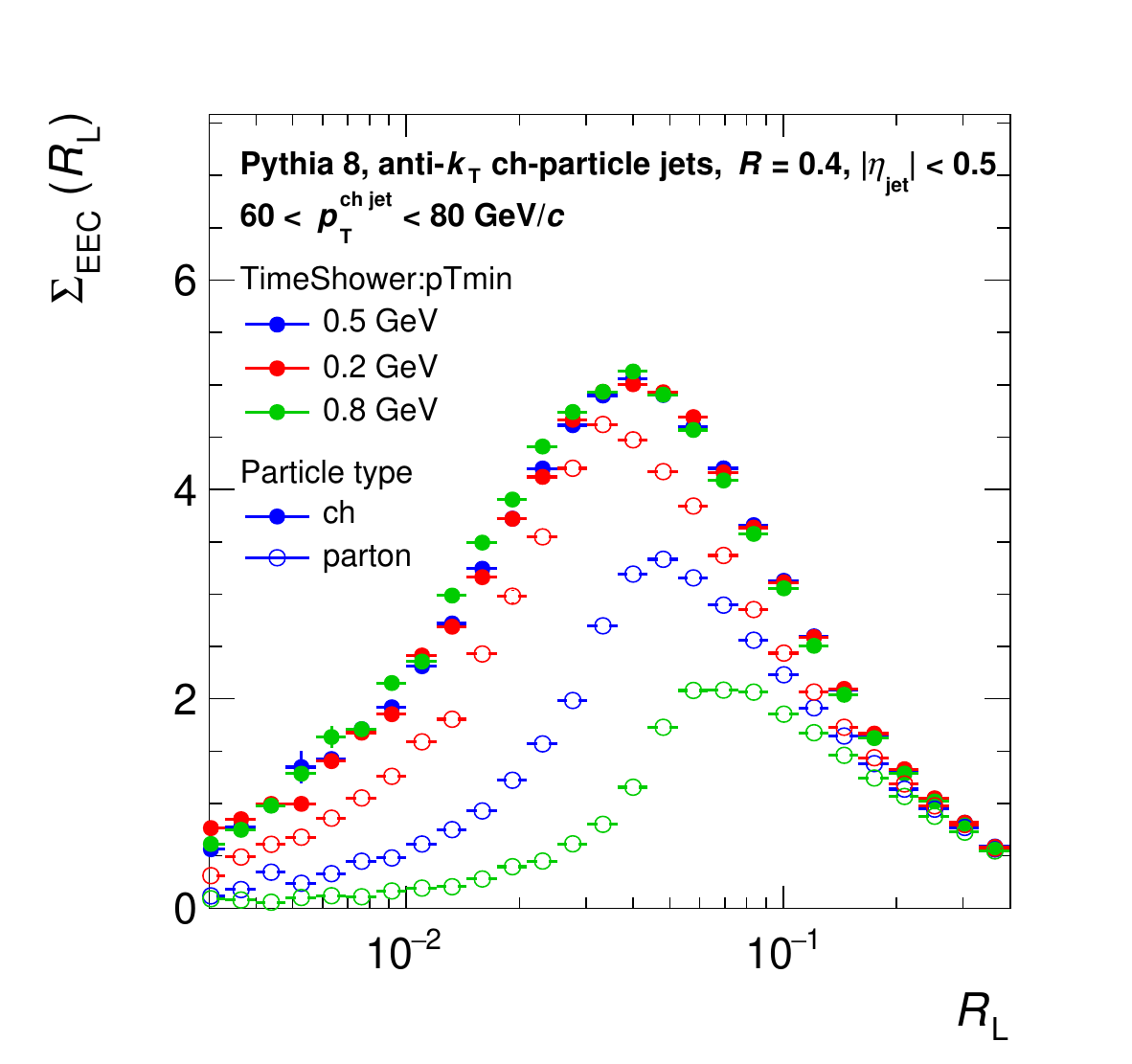}
\caption{Comparison of \sigmaeec~as a function of \RL~with different values for TimeShower:pTmin using generated Pythia events. The default setting is 0.5 (blue markers). The parton level distributions are made from parton jets that are matched to the charged jets with \pT~range indicated in each figure.}
\label{fig:pythia_ptmin}
\end{figure}

\begin{figure}[!htb]
\centering{}
\includegraphics[scale=0.39]{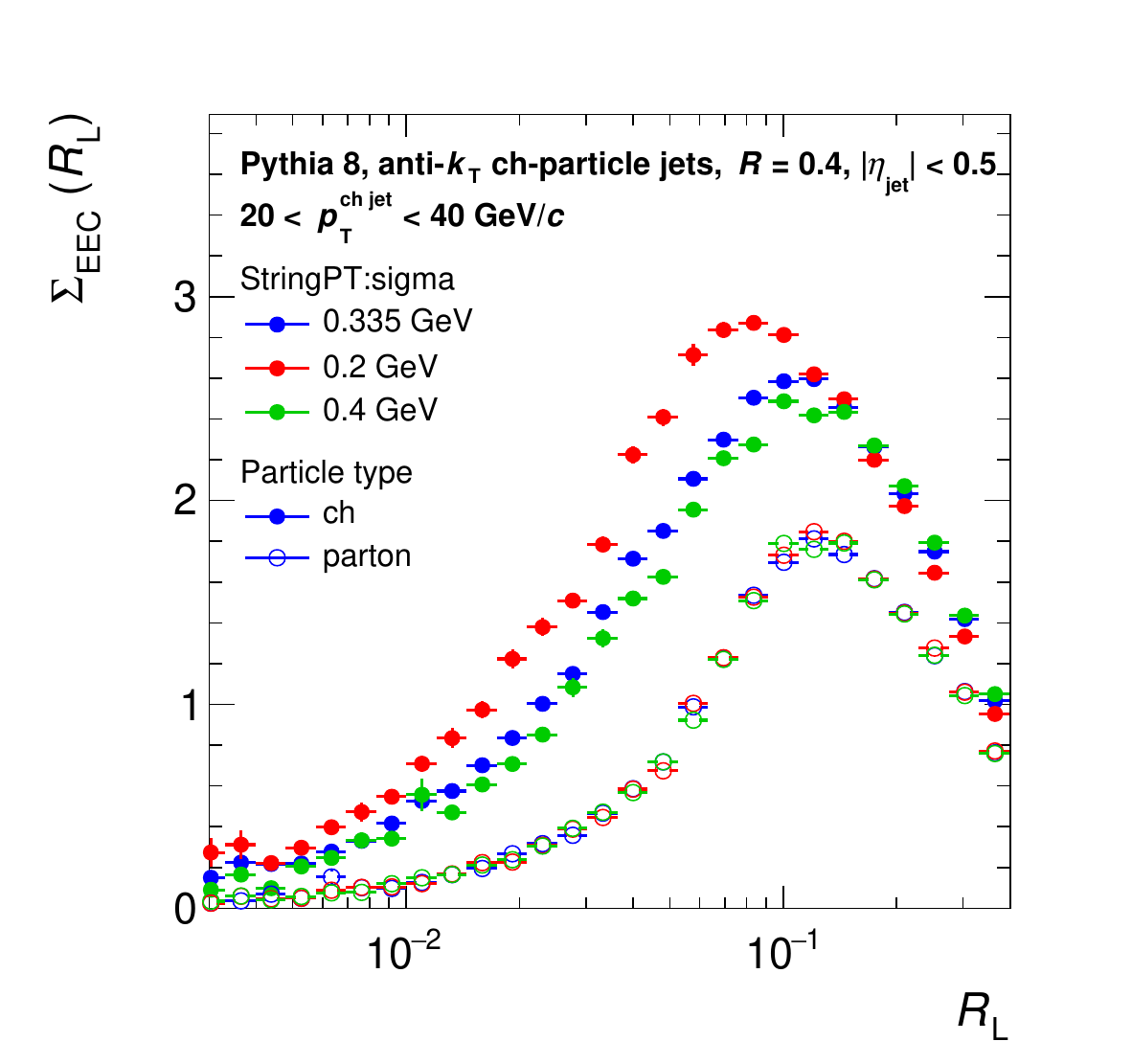}
\includegraphics[scale=0.39]{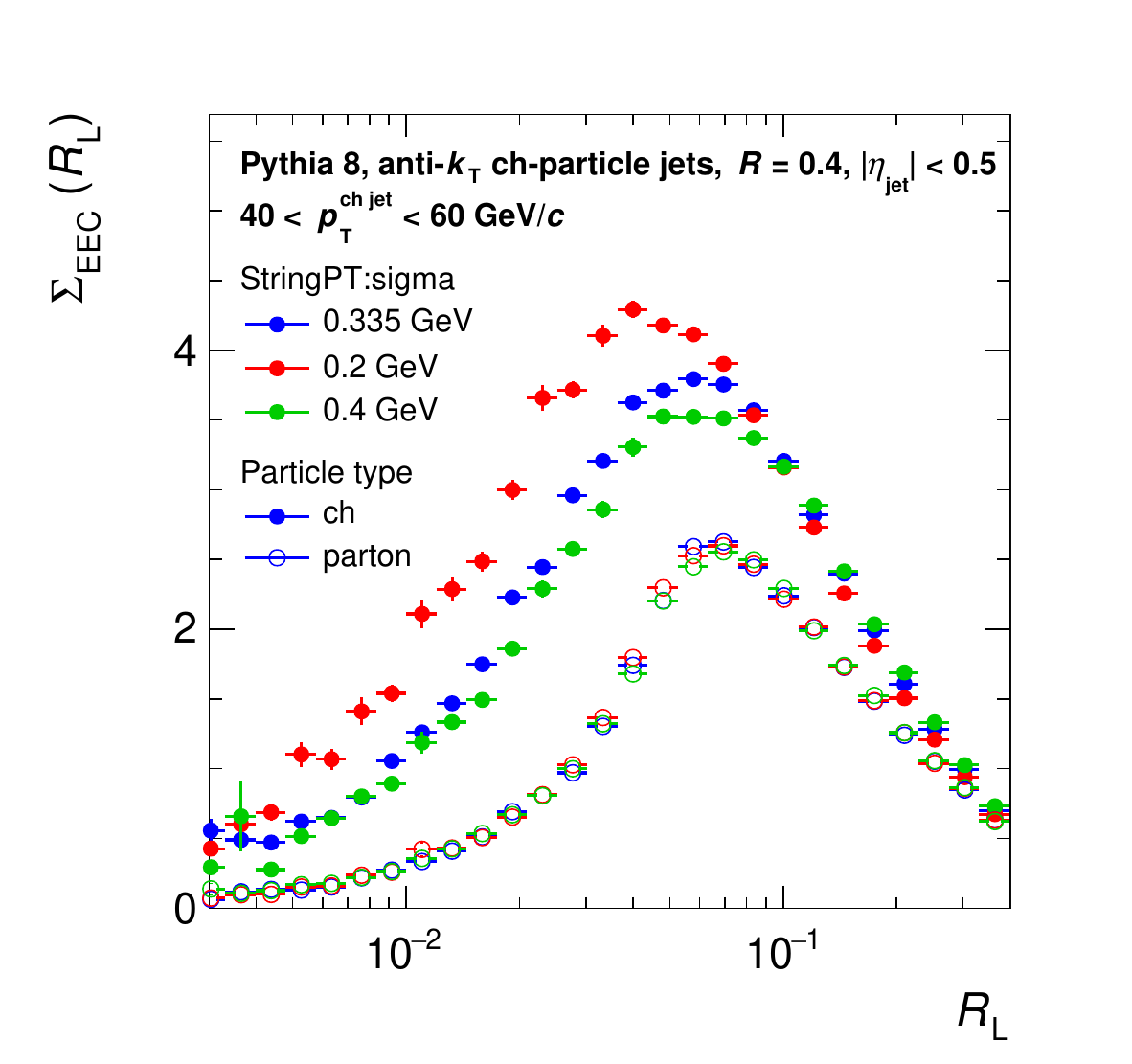}
\includegraphics[scale=0.39]{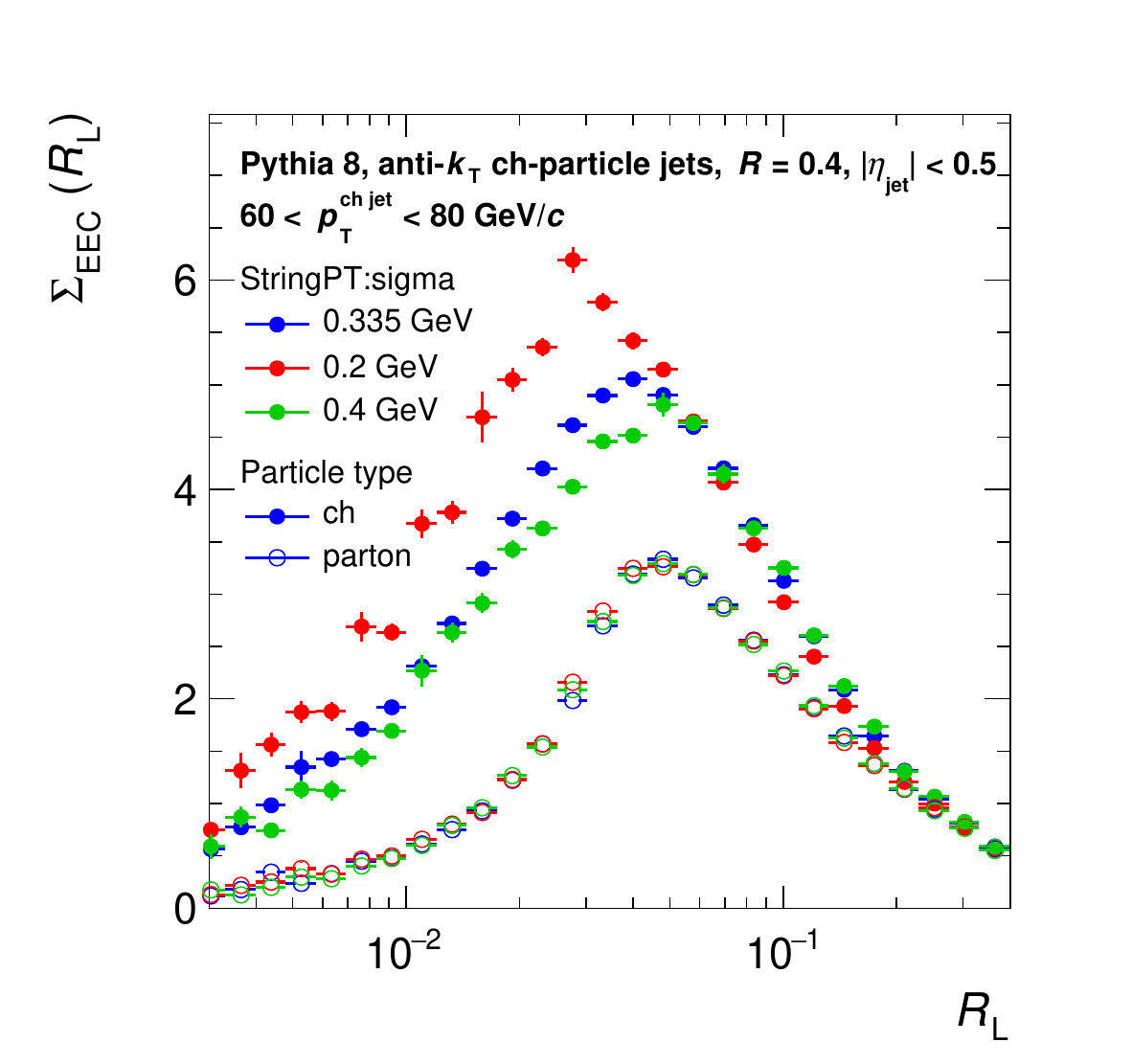}
\caption{Comparison of \sigmaeec~section as a function of $R_L$ with different values for StringPT:sigma using generated PYTHIA 8 events. The default setting is 0.335 (blue markers). The parton level distributions are made from parton jets that are matched to the charged jets with \pT~range indicated in each figure.}
\label{fig:pythia_ptsigma}
\end{figure}

We then vary the above parameters up and down by $30\%$ from the default values and compare how it impacts the EEC cross section as function of \RL. We looked both at the charged hadron level and parton level to gain further insight into the impact at different stages of the jet evolution modeled by PYTHIA 8. The results are shown in Fig.~\ref{fig:pythia_ptmin} and Fig.~\ref{fig:pythia_ptsigma}. The blue markers always correspond to the default PYTHIA 8 tuning in Monash tune. 

From Fig.~\ref{fig:pythia_ptmin}, it can be seen that the TimeShower:pTmin parameter has a significant impact on the EECs at the parton level but shows almost no visible impact on the EEC at the charged hadron level. Such an observation indicates that the transition peak of the hadron level EEC (what can be measured) is not very sensitive to the QCD shower cutoff parameter. Despite not being able to measure the parton level EEC, it is interesting to see that there is a turnover transition region and the location of that moves to a smaller angle when changing the QCD shower cutoff to a smaller value, while the transition location moves towards a larger angle with increasing QCD shower cutoff. This is consistent with the intuition that the \RL~is related to the energy scale of the parton splitting, hence when we lower the shower cutoff, the EEC continues on the parton shower trend for a wider \RL~region. Once the hadronization effects are included, the transition region with different TimeShower:pTmin appear at a very similar location.

Contrary to the observation in Fig.~\ref{fig:pythia_ptmin}, in Fig.~\ref{fig:pythia_ptsigma} we observe that when we alter StringPT:sigma, which is directly related to the hadronization process, a significant change is observed in the EECs at hadron level but no visible difference at the parton level. Since StringPT:sigma is only related to the hadron fragmentation, it makes sense that the variation of this parameter does not affect the EECs at parton level. On the other hand, at the hadron level, the transition peak seems to be sensitive to the variations of StringPT:sigma. The transition location shifts to smaller angles when we decrease this parameter. The smaller StringPT:sigma corresponds to smaller string tension where the Lund strings break up later and a lower energy scale. This picture is consistent with the observed shift when StringPT:sigma is decreased.

These PYTHIA 8 studies support that the transition peak is sensitive to hadronization process. We hope these studies and the reported measurement can motivate first principle calculations which can help draw conclusive statements. 
\clearpage

\clearpage

\section{The ALICE Collaboration}
\label{app:collab}
\begin{flushleft} 
\small

S.~Acharya\,\orcidlink{0000-0002-9213-5329}\,$^{\rm 126}$, 
A.~Agarwal$^{\rm 134}$, 
G.~Aglieri Rinella\,\orcidlink{0000-0002-9611-3696}\,$^{\rm 32}$, 
L.~Aglietta\,\orcidlink{0009-0003-0763-6802}\,$^{\rm 24}$, 
M.~Agnello\,\orcidlink{0000-0002-0760-5075}\,$^{\rm 29}$, 
N.~Agrawal\,\orcidlink{0000-0003-0348-9836}\,$^{\rm 25}$, 
Z.~Ahammed\,\orcidlink{0000-0001-5241-7412}\,$^{\rm 134}$, 
S.~Ahmad\,\orcidlink{0000-0003-0497-5705}\,$^{\rm 15}$, 
S.U.~Ahn\,\orcidlink{0000-0001-8847-489X}\,$^{\rm 71}$, 
I.~Ahuja\,\orcidlink{0000-0002-4417-1392}\,$^{\rm 36}$
% ...

\section*{Affiliation Notes}

$^{\rm I}$ Deceased\\
$^{\rm II}$ Also at: Max-Planck-Institut fur Physik, Munich, Germany\\
$^{\rm III}$ Also at: Italian National Agency for New Technologies, Energy and Sustainable Economic Development (ENEA), Bologna, Italy\\
$^{\rm IV}$ Also at: Dipartimento DET del Politecnico di Torino, Turin, Italy\\
$^{\rm V}$ Also at: Department of Applied Physics, Aligarh Muslim University, Aligarh, India\\
$^{\rm VI}$ Also at: Institute of Theoretical Physics, University of Wroclaw, Poland\\
$^{\rm VII}$ Also at: Facultad de Ciencias, Universidad Nacional Autónoma de México, Mexico City, Mexico\\

\section*{Collaboration Institutes}

$^{1}$ A.I. Alikhanyan National Science Laboratory (Yerevan Physics Institute) Foundation, Yerevan, Armenia\\
$^{2}$ AGH University of Krakow, Cracow, Poland\\
$^{3}$ Bogolyubov Institute for Theoretical Physics, National Academy of Sciences of Ukraine, Kiev, Ukraine\\
$^{4}$ Bose Institute, Department of Physics  and Centre for Astroparticle Physics and Space Science (CAPSS), Kolkata, India\\
$^{5}$ California Polytechnic State University, San Luis Obispo, California, United States\\
$^{6}$ Central China Normal University, Wuhan, China\\
$^{7}$ Centro de Aplicaciones Tecnol\'{o}gicas y Desarrollo Nuclear (CEADEN), Havana, Cuba\\
$^{8}$ Centro de Investigaci\'{o}n y de Estudios Avanzados (CINVESTAV), Mexico City and M\'{e}rida, Mexico\\
$^{9}$ Chicago State University, Chicago, Illinois, United States\\
$^{10}$ China Institute of Atomic Energy, Beijing, China\\
$^{11}$ China University of Geosciences, Wuhan, China\\
$^{12}$ Chungbuk National University, Cheongju, Republic of Korea\\
$^{13}$ Comenius University Bratislava, Faculty of Mathematics, Physics and Informatics, Bratislava, Slovak Republic\\
$^{14}$ Creighton University, Omaha, Nebraska, United States\\
$^{15}$ Department of Physics, Aligarh Muslim University, Aligarh, India\\
$^{16}$ Department of Physics, Pusan National University, Pusan, Republic of Korea\\
$^{17}$ Department of Physics, Sejong University, Seoul, Republic of Korea\\
$^{18}$ Department of Physics, University of California, Berkeley, California, United States\\
$^{19}$ Department of Physics, University of Oslo, Oslo, Norway\\
$^{20}$ Department of Physics and Technology, University of Bergen, Bergen, Norway\\
$^{21}$ Dipartimento di Fisica, Universit\`{a} di Pavia, Pavia, Italy\\
$^{22}$ Dipartimento di Fisica dell'Universit\`{a} and Sezione INFN, Cagliari, Italy\\
$^{23}$ Dipartimento di Fisica dell'Universit\`{a} and Sezione INFN, Trieste, Italy\\
$^{24}$ Dipartimento di Fisica dell'Universit\`{a} and Sezione INFN, Turin, Italy\\
$^{25}$ Dipartimento di Fisica e Astronomia dell'Universit\`{a} and Sezione INFN, Bologna, Italy\\
$^{26}$ Dipartimento di Fisica e Astronomia dell'Universit\`{a} and Sezione INFN, Catania, Italy\\
$^{27}$ Dipartimento di Fisica e Astronomia dell'Universit\`{a} and Sezione INFN, Padova, Italy\\
$^{28}$ Dipartimento di Fisica `E.R.~Caianiello' dell'Universit\`{a} and Gruppo Collegato INFN, Salerno, Italy\\
$^{29}$ Dipartimento DISAT del Politecnico and Sezione INFN, Turin, Italy\\
$^{30}$ Dipartimento di Scienze MIFT, Universit\`{a} di Messina, Messina, Italy\\
$^{31}$ Dipartimento Interateneo di Fisica `M.~Merlin' and Sezione INFN, Bari, Italy\\
$^{32}$ European Organization for Nuclear Research (CERN), Geneva, Switzerland\\
$^{33}$ Faculty of Electrical Engineering, Mechanical Engineering and Naval Architecture, University of Split, Split, Croatia\\
$^{34}$ Faculty of Nuclear Sciences and Physical Engineering, Czech Technical University in Prague, Prague, Czech Republic\\
$^{35}$ Faculty of Physics, Sofia University, Sofia, Bulgaria\\
$^{36}$ Faculty of Science, P.J.~\v{S}af\'{a}rik University, Ko\v{s}ice, Slovak Republic\\
$^{37}$ Faculty of Technology, Environmental and Social Sciences, Bergen, Norway\\
$^{38}$ Frankfurt Institute for Advanced Studies, Johann Wolfgang Goethe-Universit\"{a}t Frankfurt, Frankfurt, Germany\\
$^{39}$ Fudan University, Shanghai, China\\
$^{40}$ Gangneung-Wonju National University, Gangneung, Republic of Korea\\
$^{41}$ Gauhati University, Department of Physics, Guwahati, India\\
$^{42}$ Helmholtz-Institut f\"{u}r Strahlen- und Kernphysik, Rheinische Friedrich-Wilhelms-Universit\"{a}t Bonn, Bonn, Germany\\
$^{43}$ Helsinki Institute of Physics (HIP), Helsinki, Finland\\
$^{44}$ High Energy Physics Group,  Universidad Aut\'{o}noma de Puebla, Puebla, Mexico\\
$^{45}$ Horia Hulubei National Institute of Physics and Nuclear Engineering, Bucharest, Romania\\
$^{46}$ HUN-REN Wigner Research Centre for Physics, Budapest, Hungary\\
$^{47}$ Indian Institute of Technology Bombay (IIT), Mumbai, India\\
$^{48}$ Indian Institute of Technology Indore, Indore, India\\
$^{49}$ INFN, Laboratori Nazionali di Frascati, Frascati, Italy\\
$^{50}$ INFN, Sezione di Bari, Bari, Italy\\
$^{51}$ INFN, Sezione di Bologna, Bologna, Italy\\
$^{52}$ INFN, Sezione di Cagliari, Cagliari, Italy\\
$^{53}$ INFN, Sezione di Catania, Catania, Italy\\
$^{54}$ INFN, Sezione di Padova, Padova, Italy\\
$^{55}$ INFN, Sezione di Pavia, Pavia, Italy\\
$^{56}$ INFN, Sezione di Torino, Turin, Italy\\
$^{57}$ INFN, Sezione di Trieste, Trieste, Italy\\
$^{58}$ Inha University, Incheon, Republic of Korea\\
$^{59}$ Institute for Gravitational and Subatomic Physics (GRASP), Utrecht University/Nikhef, Utrecht, Netherlands\\
$^{60}$ Institute of Experimental Physics, Slovak Academy of Sciences, Ko\v{s}ice, Slovak Republic\\
$^{61}$ Institute of Physics, Homi Bhabha National Institute, Bhubaneswar, India\\
$^{62}$ Institute of Physics of the Czech Academy of Sciences, Prague, Czech Republic\\
$^{63}$ Institute of Space Science (ISS), Bucharest, Romania\\
$^{64}$ Institut f\"{u}r Kernphysik, Johann Wolfgang Goethe-Universit\"{a}t Frankfurt, Frankfurt, Germany\\
$^{65}$ Instituto de Ciencias Nucleares, Universidad Nacional Aut\'{o}noma de M\'{e}xico, Mexico City, Mexico\\
$^{66}$ Instituto de F\'{i}sica, Universidade Federal do Rio Grande do Sul (UFRGS), Porto Alegre, Brazil\\
$^{67}$ Instituto de F\'{\i}sica, Universidad Nacional Aut\'{o}noma de M\'{e}xico, Mexico City, Mexico\\
$^{68}$ iThemba LABS, National Research Foundation, Somerset West, South Africa\\
$^{69}$ Jeonbuk National University, Jeonju, Republic of Korea\\
$^{70}$ Johann-Wolfgang-Goethe Universit\"{a}t Frankfurt Institut f\"{u}r Informatik, Fachbereich Informatik und Mathematik, Frankfurt, Germany\\
$^{71}$ Korea Institute of Science and Technology Information, Daejeon, Republic of Korea\\
$^{72}$ Laboratoire de Physique Subatomique et de Cosmologie, Universit\'{e} Grenoble-Alpes, CNRS-IN2P3, Grenoble, France\\
$^{73}$ Lawrence Berkeley National Laboratory, Berkeley, California, United States\\
$^{74}$ Lund University Department of Physics, Division of Particle Physics, Lund, Sweden\\
$^{75}$ Nagasaki Institute of Applied Science, Nagasaki, Japan\\
$^{76}$ Nara Women{'}s University (NWU), Nara, Japan\\
$^{77}$ National and Kapodistrian University of Athens, School of Science, Department of Physics , Athens, Greece\\
$^{78}$ National Centre for Nuclear Research, Warsaw, Poland\\
$^{79}$ National Institute of Science Education and Research, Homi Bhabha National Institute, Jatni, India\\
$^{80}$ National Nuclear Research Center, Baku, Azerbaijan\\
$^{81}$ National Research and Innovation Agency - BRIN, Jakarta, Indonesia\\
$^{82}$ Niels Bohr Institute, University of Copenhagen, Copenhagen, Denmark\\
$^{83}$ Nikhef, National institute for subatomic physics, Amsterdam, Netherlands\\
$^{84}$ Nuclear Physics Group, STFC Daresbury Laboratory, Daresbury, United Kingdom\\
$^{85}$ Nuclear Physics Institute of the Czech Academy of Sciences, Husinec-\v{R}e\v{z}, Czech Republic\\
$^{86}$ Oak Ridge National Laboratory, Oak Ridge, Tennessee, United States\\
$^{87}$ Ohio State University, Columbus, Ohio, United States\\
$^{88}$ Physics department, Faculty of science, University of Zagreb, Zagreb, Croatia\\
$^{89}$ Physics Department, Panjab University, Chandigarh, India\\
$^{90}$ Physics Department, University of Jammu, Jammu, India\\
$^{91}$ Physics Program and International Institute for Sustainability with Knotted Chiral Meta Matter (WPI-SKCM$^{2}$), Hiroshima University, Hiroshima, Japan\\
$^{92}$ Physikalisches Institut, Eberhard-Karls-Universit\"{a}t T\"{u}bingen, T\"{u}bingen, Germany\\
$^{93}$ Physikalisches Institut, Ruprecht-Karls-Universit\"{a}t Heidelberg, Heidelberg, Germany\\
$^{94}$ Physik Department, Technische Universit\"{a}t M\"{u}nchen, Munich, Germany\\
$^{95}$ Politecnico di Bari and Sezione INFN, Bari, Italy\\
$^{96}$ Research Division and ExtreMe Matter Institute EMMI, GSI Helmholtzzentrum f\"ur Schwerionenforschung GmbH, Darmstadt, Germany\\
$^{97}$ Saga University, Saga, Japan\\
$^{98}$ Saha Institute of Nuclear Physics, Homi Bhabha National Institute, Kolkata, India\\
$^{99}$ School of Physics and Astronomy, University of Birmingham, Birmingham, United Kingdom\\
$^{100}$ Secci\'{o}n F\'{\i}sica, Departamento de Ciencias, Pontificia Universidad Cat\'{o}lica del Per\'{u}, Lima, Peru\\
$^{101}$ Stefan Meyer Institut f\"{u}r Subatomare Physik (SMI), Vienna, Austria\\
$^{102}$ SUBATECH, IMT Atlantique, Nantes Universit\'{e}, CNRS-IN2P3, Nantes, France\\
$^{103}$ Sungkyunkwan University, Suwon City, Republic of Korea\\
$^{104}$ Suranaree University of Technology, Nakhon Ratchasima, Thailand\\
$^{105}$ Technical University of Ko\v{s}ice, Ko\v{s}ice, Slovak Republic\\
$^{106}$ The Henryk Niewodniczanski Institute of Nuclear Physics, Polish Academy of Sciences, Cracow, Poland\\
$^{107}$ The University of Texas at Austin, Austin, Texas, United States\\
$^{108}$ Universidad Aut\'{o}noma de Sinaloa, Culiac\'{a}n, Mexico\\
$^{109}$ Universidade de S\~{a}o Paulo (USP), S\~{a}o Paulo, Brazil\\
$^{110}$ Universidade Estadual de Campinas (UNICAMP), Campinas, Brazil\\
$^{111}$ Universidade Federal do ABC, Santo Andre, Brazil\\
$^{112}$ Universitatea Nationala de Stiinta si Tehnologie Politehnica Bucuresti, Bucharest, Romania\\
$^{113}$ University of Cape Town, Cape Town, South Africa\\
$^{114}$ University of Derby, Derby, United Kingdom\\
$^{115}$ University of Houston, Houston, Texas, United States\\
$^{116}$ University of Jyv\"{a}skyl\"{a}, Jyv\"{a}skyl\"{a}, Finland\\
$^{117}$ University of Kansas, Lawrence, Kansas, United States\\
$^{118}$ University of Liverpool, Liverpool, United Kingdom\\
$^{119}$ University of Science and Technology of China, Hefei, China\\
$^{120}$ University of South-Eastern Norway, Kongsberg, Norway\\
$^{121}$ University of Tennessee, Knoxville, Tennessee, United States\\
$^{122}$ University of the Witwatersrand, Johannesburg, South Africa\\
$^{123}$ University of Tokyo, Tokyo, Japan\\
$^{124}$ University of Tsukuba, Tsukuba, Japan\\
$^{125}$ Universit\"{a}t M\"{u}nster, Institut f\"{u}r Kernphysik, M\"{u}nster, Germany\\
$^{126}$ Universit\'{e} Clermont Auvergne, CNRS/IN2P3, LPC, Clermont-Ferrand, France\\
$^{127}$ Universit\'{e} de Lyon, CNRS/IN2P3, Institut de Physique des 2 Infinis de Lyon, Lyon, France\\
$^{128}$ Universit\'{e} de Strasbourg, CNRS, IPHC UMR 7178, F-67000 Strasbourg, France, Strasbourg, France\\
$^{129}$ Universit\'{e} Paris-Saclay, Centre d'Etudes de Saclay (CEA), IRFU, D\'{e}partment de Physique Nucl\'{e}aire (DPhN), Saclay, France\\
$^{130}$ Universit\'{e}  Paris-Saclay, CNRS/IN2P3, IJCLab, Orsay, France\\
$^{131}$ Universit\`{a} degli Studi di Foggia, Foggia, Italy\\
$^{132}$ Universit\`{a} del Piemonte Orientale, Vercelli, Italy\\
$^{133}$ Universit\`{a} di Brescia, Brescia, Italy\\
$^{134}$ Variable Energy Cyclotron Centre, Homi Bhabha National Institute, Kolkata, India\\
$^{135}$ Warsaw University of Technology, Warsaw, Poland\\
$^{136}$ Wayne State University, Detroit, Michigan, United States\\
$^{137}$ Yale University, New Haven, Connecticut, United States\\
$^{138}$ Yildiz Technical University, Istanbul, Turkey\\
$^{139}$ Yonsei University, Seoul, Republic of Korea\\
$^{140}$ Affiliated with an institute covered by a cooperation agreement with CERN\\
$^{141}$ Affiliated with an international laboratory covered by a cooperation agreement with CERN.\\

\end{flushleft} 

\end{document}